\documentclass{aa}  

\usepackage{graphicx}
\usepackage{txfonts}
\usepackage{lscape}
\newcommand{\Gaia}{{\it Gaia}}
\usepackage{hyperref}
\hypersetup{colorlinks=true, citecolor=blue}
\begin{document}

   \title{Characterisation of Galactic carbon stars  and related stars from \Gaia\  EDR3}

   \subtitle{}

  \author{C. Abia\inst{1}
  %\orcidlink{0000-0002-5665-2716}
          \and
           P. de Laverny\inst{2}%\orcidlink{0000-0002-2817-4104}
          \and
          M. Romero-G\'omez\inst{3}
          \and
          F. Figueras\inst{3}
          }

   \institute{Dpto. F\'\i sica Te\'orica y del Cosmos, Universidad de Granada,
              E-18071 Granada, Spain,
              \email{cabia@ugr.es}
              \and 
              Universit\'e C\^ote d’Azur, Observatoire de la C\^ote d’Azur, CNRS, Laboratoire Lagrange, France
              \and
              Institut de Ci\`encies del Cosmos, Universitat de Barcelona (IEEC-UB), Mart\'\i \hspace{0.05cm} i Franqu\`es 1, 08028 Barcelona, Spain
          }

   \date{Received ; accepted }

% \abstract{}{}{}{}{} 
% 5 {} token are mandatory
 
  \abstract {The third  early \Gaia\ data release  (EDR3) has improved
    the accuracy of the astrometric parameters of numerous long-period
    variable (LPV)  stars. Many of  these stars are on  the asymptotic
    giant branch (AGB), showing either a C-rich or O-rich envelope and
    are   characterised   by   high   luminosity,   changing   surface
    composition, and intense mass loss. This make them very useful for
    stellar studies. In  a previous investigation, we  used \Gaia\ DR2
    astrometry   to   derive   the  luminosity   function,   kinematic
    properties, and  stellar population  membership of  a flux-limited
    sample of  carbon stars  in the  solar neighbourhood  of different
    spectral  types.  Here,  we  extend this  initial  study  to  more
    recent surveys  with  a greater number of Galactic  carbon stars and  related stars by adopting  the more
    accurate  EDR3 astrometry  measurements.  Based on  a much  larger
    statistics, we  confirm that N-  and SC-type carbon stars  share a
    very similar luminosity function, while the luminosities of J-type
    stars  (M$_{\rm{bol}}$)  are  fainter   by  half  a  magnitude  on
    average. R-hot type carbon  stars have luminosities throughout the
    RGB, which favours the hypothesis  of an external origin for their
    carbon  enhancement.  Moreover,  the  kinematic  properties  of  a
    significant fraction  of the R-hot  stars are compatible  with the
    thick-disc population, in
    contrast with  that of  N- and SC-type  stars, which  would belong
    mostly to the  thin disk.  We also derive  the luminosity function
    of a large  number of Galactic extrinsic and  intrinsic (O-rich) S
    stars and show  that the luminosities of the  latter are typically
    higher than the predicted onset  of the third dredge-up during the
    AGB for  solar metallicity. This  result is consistent  with these
    stars  being genuine  thermally pulsing  AGB stars.  On the  other
    hand, using  the so-called \Gaia-2MASS  diagram, we show  that the
    overwhelming majority of the carbon stars identified in the LAMOST
    survey   as  AGB   stars   are  probably   R-hot  and/or   CH-type
    stars. Finally, we report the  identification of $\sim 2\,660$ new
    carbon  stars candidates  that we  identified through  their 2MASS
    photometry,  their \Gaia\  astrometry, and  their location  in the
    \Gaia-2MASS diagram.}
  


   \maketitle
%
%-------------------------------------------------------------------

\section{Introduction}
In a  previous study \citep[][hereafter  Paper I]{abi20}, a  sample of
210 carbon stars of N-, SC-,  J- and R-hot spectral types was selected
in  the solar  vicinity, mainly  based on  the flux-limited  sample of
carbon stars by \citet{cla87}. The main aim of the project is to fully
characterise the  properties of  these stars  in terms  of luminosity,
kinematics, and chemical composition to distinguish their evolutionary
status  and typical  stellar masses  and to  define to  which Galactic
population they  belong. Although the  chemical properties of  each of
these  carbon  star  types  are  relatively  well  known  \citep[][and
  references therein; see  also Paper I]{wal98,abi03,hab04,zam09}, the
conclusions  deduced  from  these  abundance  studies  are  frequently
limited by the uncertain determination of the luminosity and kinematic
properties of these stars. The  \Gaia\ all-sky survey has changed this
situation by providing astrometric data with an unprecedented accuracy
of all  Galactic stellar  populations \citep{gai18}, in  particular of
long period variables (LPV) \citep{mow18}, among  which most of these carbon  star types are
found.

In Paper I, we used \Gaia\  DR2 astrometric measurements to derive the
luminosities and kinematic properties  of already well-known red giant
carbon  stars  of  different  spectral  types  located  in  the  solar
vicinity. We found that the luminosity function (LF) of N- and SC-type
stars is quite  similar, but is clearly different from  that of J-type
stars. Interestingly,  we also  found extended tails  to low  and high
luminosity  in  the   LF  of  the  N-type  stars   and  discussed  the
implications that  this may have  for the theoretical lower  and upper
mass limits  for the formation of  intrinsic AGB carbon stars.  On the
other hand,  the derived LF of  R-hot stars revealed that  these stars
should be in the RGB phase rather than  in the red clump of the HB, as
previously suggested \citep{ber02}. However, these findings were based
on a small statistics, in  particular for SC-type (10 objects), J-type
(21 objects), and R-hot stars (35 objects).

In the  present study, we  extend our  previous analysis to  more than
$2\,000$ objects that  we extracted from different  recent giant stars
surveys. All  the different spectral  types of carbon  stars mentioned
above were  considered, and we  adopted the more accurate  \Gaia\ EDR3
astrometric data \citep{lin21}. Our present analysis also includes the
identification of a large number of  new AGB carbon star candidates in
the  2MASS survey  \citep{skr06} through  the use  of the  \Gaia-2MASS
diagram \citep{leb18}.  The  sample is introduced in Sect.  2, and the
analysis is carried out  in Sects. 3 and 4. In  Sect.~5 we explore the
capability  of  the  \Gaia-2MASS  diagram to  identify  the  different
spectral  types of  AGB stars,  and we  present the  catalogue of  new
carbon star  candidates in Sect. 6.  A summary of the  main results of
this study is made in Sect. 7.

%--------------------------------------------------------------------
\section{Star sample and Galactic distribution}
\label{SecSample}
 We  extended  the sample  of  carbon  stars  studied  in Paper  I  by
 including  the collected  Galactic infrared  carbon stars  (IRCSs) in
 \citet{che12}, the new catalogue of  carbon stars from the LAMOST DR2
 survey \citep{ji16}, and the list  of J-type Galactic carbon stars in
 \citet{che07}. We  recall that  the carbon stars  studied in  Paper I
 were  taken from  the flux-limited  sample of  Galactic carbon  stars
 studied by  \citet{cla87}, which includes  about 215 carbon  stars of
 N-, J-, and SC-types (see Paper I for details of the selection).
 
 The  sample of  \citet{che12}  collects  all of  the  IRCSs from  the
 literature   \citep{cha90,gro02,che03,gup04,leb05}  and   some  known
 carbon stars  with SiC emission at  11.2 $\mu$m that were  not listed
 previously. In total,  the sample includes 974 objects,  most of them
 quoted    as     being    carbon     stars    according     to    the
 SIMBAD\footnote{http://simbad.u-strasbg.fr/simbad/sim-fid}
 database.  Nevertheless,  a  significant  number of  objects  in  the
 \citet{che12} sample have no  spectral type identification and appear
 in SIMBAD  just as 
   possible carbon stars (see below).  The LAMOST survey includes 894
 carbon stars  that were identified  by several spectral  line indices
 from low-resolution spectra. By combining several CN and C$_2$ bands,
 \citet{ji16} identified  carbon stars  of N-,  CH-, and  R-hot types,
 although several dozen objects were  also quoted as of  possible
 N-type or of  unknown carbon type. Below, we  show that most of
 these  objects with  dubious  spectral classification  in the  LAMOST
 sample are probably not genuine AGB carbon stars.  On the other hand,
 the infrared study by \citet{che07}  included 113 Galactic stars of J
 type. A few  of them were included  in the sample of  J-type stars we
 studied in  Paper I. Finally,  for comparison purposes,  we completed
 our  stellar  sample  with  the  Galactic  S-type  stars  studied  by
 \citet{che19}:  151  extrinsic, and  190  intrinsic.  We recall  that
 intrinsic S-type stars  are O-rich objects (C/O$<1$)  that follow the
 M-S-C sequence evolution in the AGB phase. They show intense Tc lines
 and      $s$-element      enhancements     in      their      spectra
 \citep{van98,van99,ute10,ute13}. The    detection   of    the
   unstable element  technetium in  the spectra of  these stars  is an
   evidence of the  in situ operation  of the s-process
   nucleosynthesis (see e.g.  \citet{stra06, kar14}).
 Extrinsic S-type  stars are  also O-rich objects  in a  binary system
 with  a  white   dwarf  (WD,  undetected)  and   also  have  enhanced
 $s$-process elements  at the  stellar surface, but  as the  result of
 mass transfer from a  former AGB star (now the WD)  to a less evolved
 companion (now the S-type star). They do not show intense Tc lines in
 their optical spectra.
 
\begin{figure}
   \centering
   \includegraphics[width=8.5 cm]{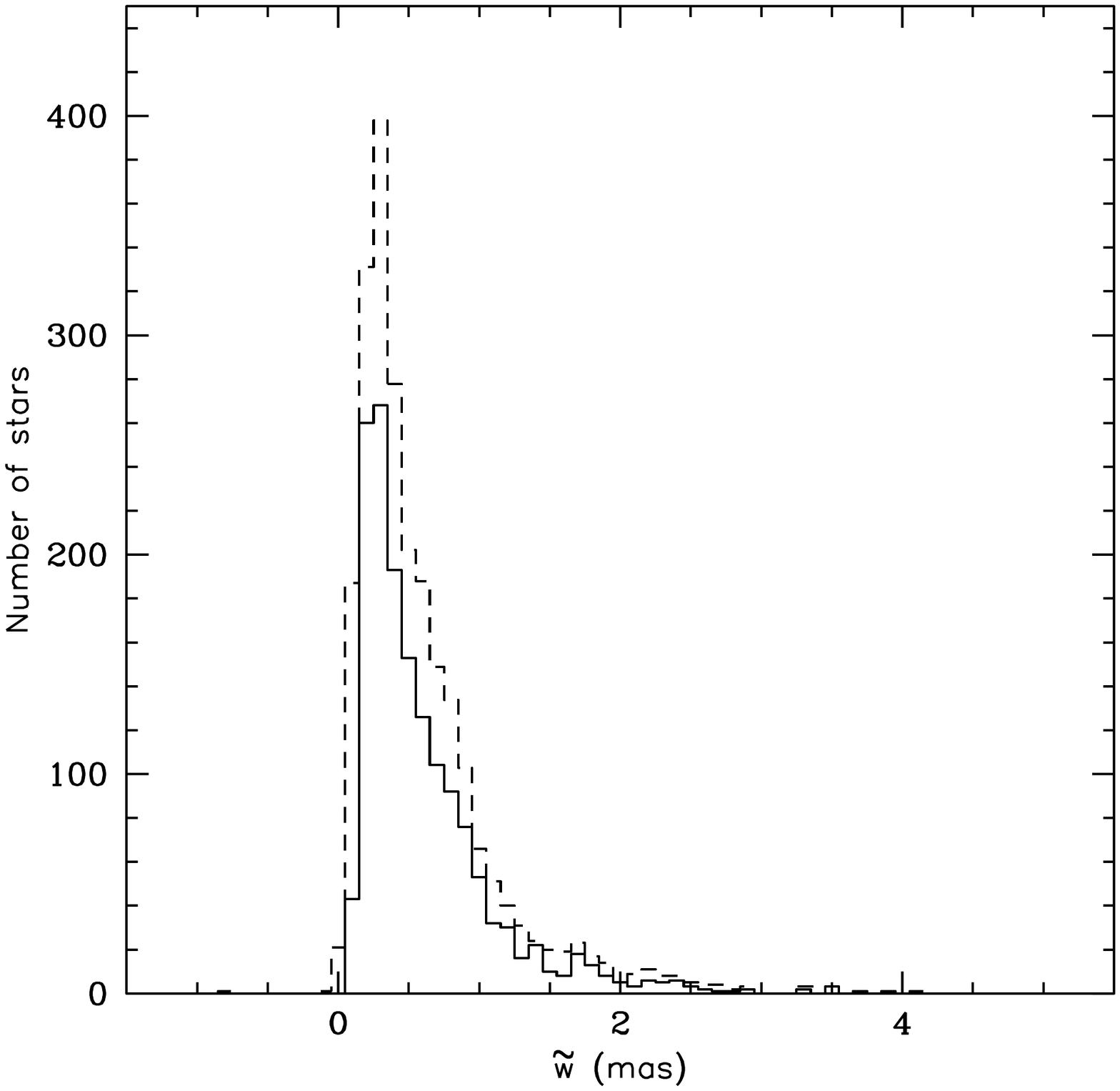}
   \caption{Comparison  of  the  histogram   of  the  {\it  Gaia}  DR3
     parallaxes in the initial sample  of stars (dashed line) with the
     parallaxes deduced from the \citet{bai21} distances adopted after
     filtering the  sample according to our  quality parallax criteria
     (continuous line). The bin size is 0.1 mas (see text).}
  \end{figure}

 Next, we cross-match  the stars of the \citet{che07}, \citet{che12}, and
 \citet{ji16}  samples with  the stars  studied  in Paper  I to  avoid
 repetitions.  We  found 82 objects  in common  with Paper I  that are
 distributed among all spectral types of the carbon stars. These stars
 were removed from the surveys mentioned  above, and this was then our
 initial sample of stars. In contrast to the stars studied in Paper I,
 there  is  not  information  in the  literature  about  the  detailed
 chemical composition for the overwhelming  majority of the new carbon
 stars  included  here. However,  considering  their  position in  the
 Galactic plane and  their kinematics (see below and  Section 4), most
 of  them belong  to the  Galactic thin-disc  population. We  thus may
 guess that their metallicity is  close to solar. Therefore and unless
 explicitly  mentioned, we  assume solar  metallicity for  all of  the
 stars in our sample throughout this paper. This does not apply to the
 CH-type stars in the LAMOST sample;  most of these stars are known to
 be metal-poor objects \citep[see e.g.][]{wal98}. Finally, we filtered
 this sample  according to the  accuracy of their EDR3  parallaxes and
 the corresponding derived distances as described below.

\begin{figure}
   \centering
   \includegraphics[width=8.5cm]{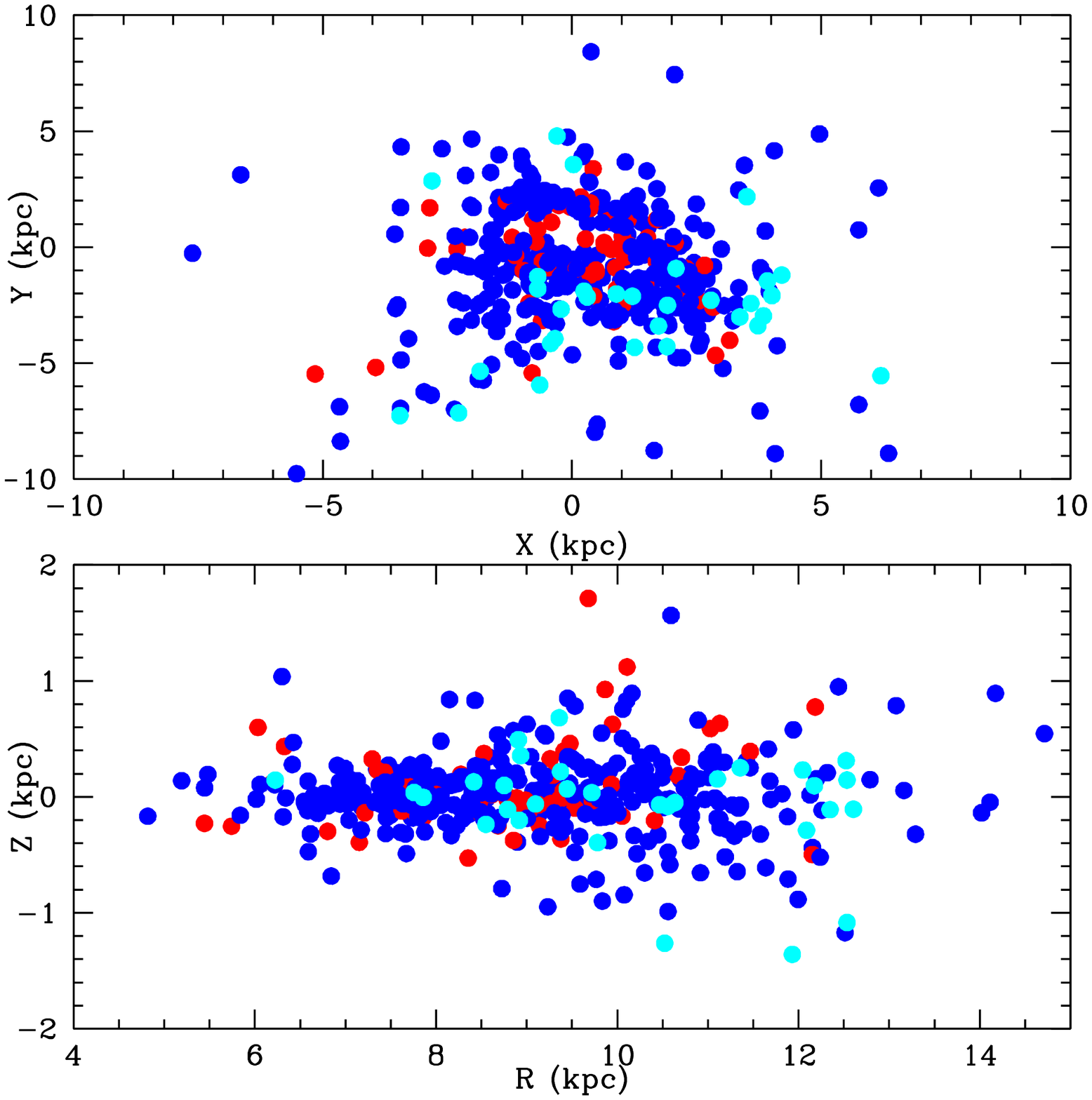}
   \caption{Top: Location of the stellar sample of carbon stars around
     the Sun. The Sun is placed at (X,Y)$=(0,0)$. Stars are taken from
     the surveys  by \citet{che07}  and \citet{che12}. Blue  dots show
     normal N-type carbon stars; red  dots represent J-type stars, and
     cyan  dots  show  stars  of   {\it  unknown}  type  according  to
     \citet{che12} that  are labelled  {\it possible} carbon  stars in
     SIMBAD. Bottom:  Distribution above  or below the  Galactic plane
     vs.  galactocentric distance.  The Sun  is at  R$=8.34$ kpc.  The
     typical uncertainty in the (X, Y,  Z) coordinates is $\pm 20$ pc,
     and  it is  $\pm  50$  pc for  the  galactocentric distance  (see
     text).}
  \end{figure}

To  recover the  most  accurate  distances of  these  stars, we  first
selected  those   with  \Gaia\  parallaxes  that   were  derived  from
five-parameter astrometric  solutions with a renormalised  unit weight
error,  RUWE, smaller  than 1.4.   We then  selected those  whose 
  astrometric fidelity factor for their astrometric solution exceeded
0.5,  according  to  \citet{Ryb22}.  This  factor  allows  identifying
possible  spurious   parallaxes  and   the  best   \Gaia\  astrometric
solutions.    For   the   selected   stars,  we   then   adopted   the
photo-geometric distances from \citet{bai21}.  Then, we again filtered
the  sample  and  chose  only   stars  with  a  \Gaia\  EDR3  parallax
uncertainty $\leq 10\%$. This condition ensures that the distance adopted from \citet{bai21} is
close  to  the inverse  of  the  parallax,  and therefore  has  little
dependence on  the adopted prior. In  Paper I we discussed  that large
colour  variations  (most of  our  stars  are variables  of  different
variability  types)   may  not   significantly  affect   the  parallax
measurements. We refer to the  discussion there for details.  Figure 1
compares  the  histogram of  the  {\it  Gaia}  DR3 parallaxes  in  the
original  sample of  stars with  that deduced  from the  \citet{bai21}
photo-geometric  distances adopted  in this  study after  applying our
parallax quality criteria. These  criteria include the use
  of positive  parallaxes alone. The  bin size in the  histograms is
0.1  mas, which  is the  typical  uncertainty in  the parallax  ($\sim
10\%$) for most of  the stars in the final sample.  The stars that did
not  fulfil our  parallax criteria  are  clearly removed  in the  full
parallax range. As a consequence,  the final sample should not contain
any bias on the distances of the stars.

We compared the distances obtained for  the stars in common with Paper
I with  those derived  here. We  find an  excellent agreement:  a mean
difference  of   $-34\pm  190$  pc   in  the  sense  DR2   minus  EDR3
distances. Only for two N-type stars,  IRC +60393 and V497 Pup, is the
difference  significant,  which  is  due to  the  $\sim  50\%$  higher
parallax in \Gaia\ DR2 for these  stars. For three stars (V437 Aql, RU
Vir, HIP 44812, and HIP 66317),  we find differences of $\sim 20\%$ in
the parallax  DR2-EDR3. On the  other hand,  we also found  a moderate
change  in  the EDR3  \Gaia\  photometry  values ($G_\mathrm{BP}$  and
$G_\mathrm{RP}$) for the stars V Cyg  and TX Psc with respect to those
in DR2. This will imply a  significant change in their position in the
\Gaia-2MASS diagram (see  below). In Paper I we have  reported that TX
Psc was suspected  to be a binary  star and that it  was the brightest
N-type star in our sample ($G=3.72$ mag according to DR2). This bright
$G$-magnitude may be affected by  saturation problems that could occur
for  the  brightest \Gaia\  DR2  targets  and  might also  affect  the
determination of the  parallax. In EDR3, the  estimated uncertainty in
the  $G$, $G_\mathrm{BP}$  , and  $G_\mathrm{RP}$ magnitudes  has been
considerably reduced for this star.

After the quality  parallax and distance filters  described above, our
final sample contains 491 carbon stars of N type, 22 of SC type, 83 of
J type, 234 of R-hot type, and  276 of CH type. For the spectral types
N, SC, J, and R-hot, this represents an increase of more than a factor
of two in the number of objects  with respect to Paper I. To these, we
added 107  extrinsic and  91 intrinsic S  stars of  \citet{ji16}.  The
total is then $1\,304$ objects  whose luminosity function and Galactic
location and kinematics we studied. However, some stars that fulfilled
our  parallax  and  distance  filters were  still  excluded  from  the
luminosity function analysis because of their uncertain classification
type and bolometric correction: 192 objects labelled extreme infrared,
and 37  objects as  possible carbon  stars in  the \citet{che12}
sample, and 86 and 69 objects labelled possible carbon stars and
of   unknown  type,  respectively, in  the  LAMOST  sample  (see
below). There are $1\,688$ objects in total.

The stars for which the luminosity function was derived are listed in Table~1. We preferentially adopted
their variable  star designation  \citep[from][]{sam04} or,  when this
was not  available, their IRAS  name \citep{neu84}, their name  in the
LAMOST  catalogue  \citep{zha12},  or  the  name  that  is  used  most
frequently in the literature according to the SIMBAD database. Table 1
also reports  their \Gaia-EDR3  identification (col. 2),  the distance
estimation  (col. 8),  and the  infrared 2MASS  photometry $(J,  K_s)$
(cols. 6 and 7) corrected for  Galactic extinction (cols. 3, 4, and 5;
see below).  According to the 2MASS survey, the average uncertainty in
the $J$ and  $K_s$ magnitudes in our stellar sample  is $0.05\pm 0.12$
and $0.07\pm 0.20$ mag, respectively. Many of the objects studied here
lack   an  accurate   or  individual   determination  of   the  radial
velocity. For the sake of homogeneity, we therefore adopted the values
given in \Gaia\ DR2 \citep{Katz2019} as the velocity along the line of
sight (V$_{\rm{rad}}$)  even though  the \Gaia/RVS spectral  domain is
not   optimal  for   deriving  V$_{\rm{rad}}$   in  cool   carbon-rich
stars. Nevertheless, for the stars in  Paper I for which more accurate
V$_{\rm{rad}}$ values in the literature  were found, we confirmed that
the \Gaia\ DR2  values do not significantly differ  from those adopted
in Paper I.

Figures 2, 3, and  4 show the location of the sample  stars in the X-Y
coordinates around the Sun and above and below the Galactic plane (see
also  Table  1)  for  the stars  in  \citet{che07}  and  \citet{che12}
(Fig. 2), stars in the LAMOST sample (Fig. 3), and the S-type stars in
\citet{che19}  (Fig. 4).  Their  Cartesian  coordinates were  directly
derived   from   the  \Gaia\   EDR3   sky   coordinates  and   adopted
distances. Median uncertainties  are $\pm12, \pm16, \pm  13$, and $\pm
13$ pc for X, Y, Z and R, respectively (see Paper I for details of the
derivation of the uncertainties).

\begin{figure}
   \centering
   \includegraphics[width=8.5 cm]{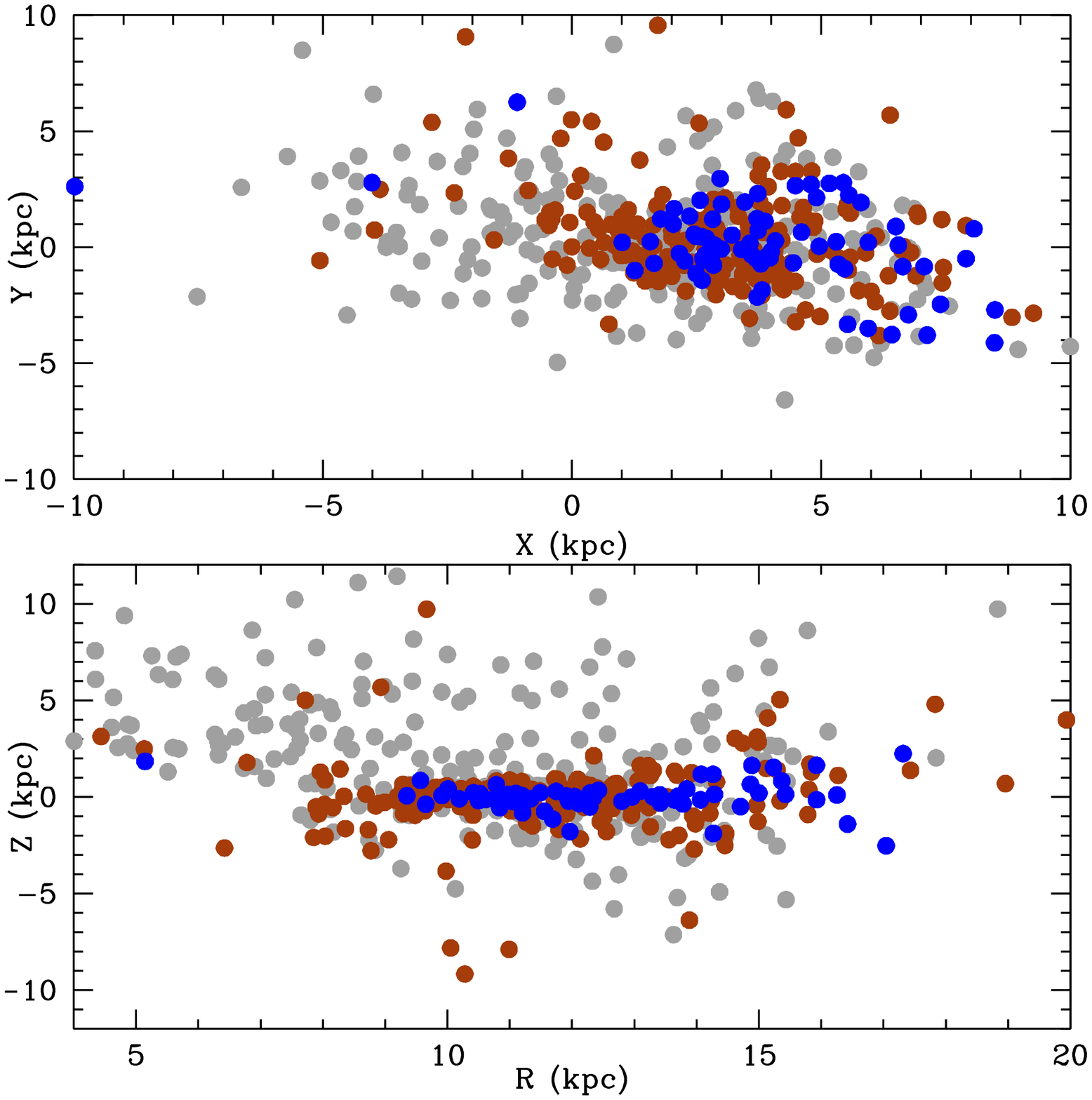}
   \caption{Same as Fig. 2 for the carbon stars in the LAMOST survey \citep{ji16}.
     Solid blue dots show N-type carbon stars,
     grey dots represent CH-type stars, and brown dots show R-hot stars.
     The limits of the X- and Y-axes and R coordinates are different from Fig. 2. }
  \end{figure}

The top panels in Figs. 2 and 4 show that the spatial distribution of the stars
in \citet{che07}, \citet{che12}, and \citet{che19} are fairly uniform around the Sun, although
the stars labelled  possible carbon stars seem to be located at Y$<0$ values and very close the Galactic
plane (cyan circles in Fig.2).

\begin{figure}
   \centering
   \includegraphics[width=9cm]{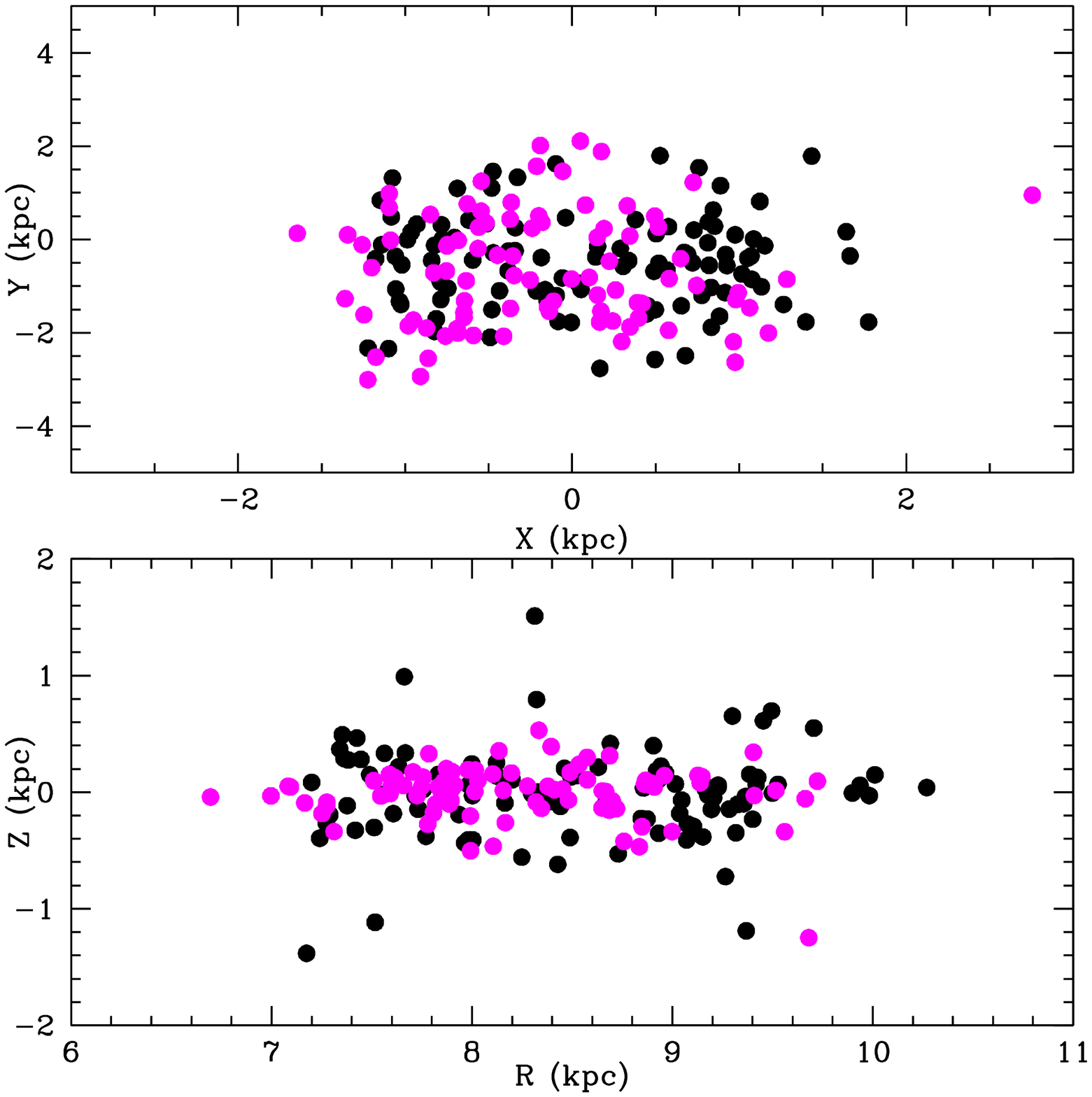}
   \caption{Same as Fig. 2 for the S stars in \citet{che19}. Solid black circles show extrinsic stars,
     and purple circles show intrinsic stars.
     The limits of the X- and Y-axes and R coordinates are different from Fig. 2.}
  \end{figure}

\begin{figure}
   \centering
   \includegraphics[width=9 cm]{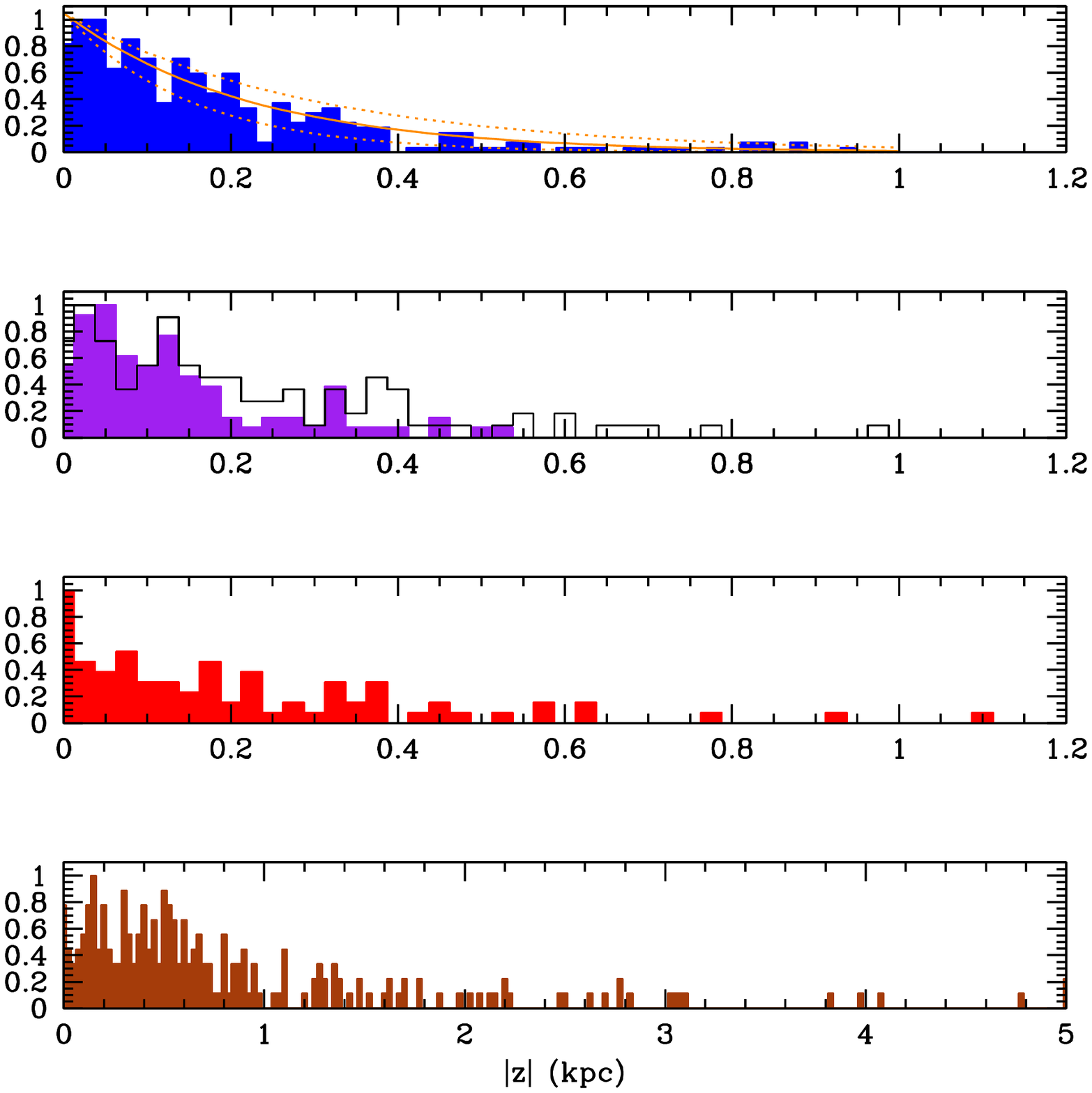}
   \caption{ Histograms showing the distribution (normalised to the maximum number of stars) of the $|Z|$-coordinate
     for the different spectral types. From top to bottom: Normal N-type carbon stars (blue), extrinsic (black) and
     intrinsic (purple) S stars, J-type stars (red), and  R-type stars (brown). The X-axis for the R-type stars is different.
     For N-type normal carbon stars, orange lines show exponential fits to the distribution with scale heights $z_o=260, 220$,
     and 180 pc from top to bottom. Similar fits give an estimate of the scale-height for the other spectral types (see text).
     The bin size is 20 pc. }
  \end{figure}

When we  consider only the  carbon stars of N  type in this  study and
those  in  Paper  I  (491  stars   in  total),  a  crude  fit  to  the
$|Z|$-coordinate distribution with an  exponential function results in
a scale  height of  $z_o\sim 220$  pc (see Fig.  5, top  panel), which
agrees   with   the   value   obtained   in   Paper   I   within   the
uncertainty. We   do   not    show   in   this   plot   the
  $|Z|$-distribution for  SC-type stars  because there  are so  few of
  them.  However, a  similar fit  is  compatible with  a scale  height
  identical to that  of N-type stars. Only $\sim 6\%$  stars of this
type  (see the  bottom panel  in  Fig. 2)  have $|Z|>  0.5$ kpc.  This
scale-height can be  used to estimate the typical mass  of carbon star
progenitors   by  using   tabulations   for  the   scale  heights   of
main-sequence    stars   as    a    function    of   spectral    class
\citep[e.g.][]{mil79}. This shows that stars  with a mean scale height
in  the   range  160-200  pc   have  a   mass  between  1.5   and  1.8
M$_\odot$.  These   values  are  fully  consistent   with  theoretical
determinations of the  typical mass for an AGB  carbon star \citep[see
  e.g.][]{stra06,kar14}. In  the same  way, while the  estimated scale
height for intrinsic S stars is  quite similar to that of N-type stars
($z_o\sim 210$  pc; see Figs. 4  and 5), extrinsic S  stars and J-type
stars  (Fig.   5)  have   higher  $z_o$  values:   280  and   300  pc,
respectively.  Because  the  scale  height  of  the  Galactic  stellar
population  is thought  to  be  a function  of  age (mass)  \citep[see
  e.g.][]{dov93},  this implies  that  these two  types  of stars  are
likely  older  and  typically  have   lower  masses  than  the  N-type
stars. This disagrees for J-type stars with our conclusion in Paper I,
where we  found a similar  scale height for  J- and N-type  stars. Our
previous conclusion  was biased by the  low number of J-type  stars in
the sample. J-type stars not  only differ in chemical composition (see
Paper  I) and  luminosity (see  below) with  respect to  normal N-type
stars, but  additionally have a different  Galactic distribution. This
might  indicate that  they are  a completely  different population  of
stars  compared to  the  N-type stars.  On the  other  hand, based  on
intrinsic S stars with a very similar scale height as N-type stars, we
may conclude  that the former  are the  progenitors of the  normal AGB
carbon stars, as expected on theoretical grounds.

The stars in the  LAMOST sample that are classified as  N type have an
identical distribution above and below the Galactic plane (see Fig. 3,
bottom panel) as those in Fig. 2.  However, we show below that many of
these stars are  too faint to be  in the AGB phase.  Fig.~3 also shows
that many R-hot carbon stars  and the overwhelming majority of CH-type
stars are located at large distances above (below) the Galactic plane;
the  average values  are  $1.2\pm 1.5$  kpc, and  $2.7  \pm 2.5$  kpc,
respectively. This is confirmed by  the |Z|-distribution of the R-type
stars (Fig. 5, bottom panel).  An exponential fit to this distribution
gives  a  scale  height  of  $z_o\sim  1.3$  kpc.  This  confirms  the
conclusion     that     was     reached    in     previous     studies
\citep{wal98,kna01,iza07,izz08,zam09} that these types of carbon stars
mainly belong  to the  Galactic thick-disc  population and  would also
have lower masses than N-type stars.

\section {Luminosities}
We retrieved the  2MASS $J$ and $K_s$ photometry  \citep{cut03} from the
SIMBAD database  and corrected them for  interstellar extinction using
the \citet{bai21}  distances and  the \Gaia\ Galactic  coordinates. We
followed the same  method as in Paper I to  derive the luminosities of
the stars.   We adopted the  empirical BC$_K$ versus  $(J-K)$ relation
for  carbon stars  obtained  by  \citet{ker10}, which  is  based on  a
critical revision  of previous  studies of bolometric  corrections for
cool giants.  For $(J-K)$ between 1.0 to 4.4 mag, the maximum standard
deviation of this relation is  0.11 mag, although for values $(J-K)\ge
2.2$ mag the  scatter increases up to 0.15 mag.  For very red objects,
in particular  for many IRAS  sources in the \citet{che12}  sample, we
therefore instead used the empirical BC$_K$ versus $K-[12.5]$ relation
from  \citet{gua06} when  the  $[12.5]$ colour  was  available in  the
literature;  otherwise, the  particular object  was rejected  from the
analysis. For  for the O-rich  S stars in  the present study,  we also
adopted the  BC$_K$ versus  $(J-K)$ relation derived  by \citet{ker10}
for M stars (see Paper I for details).

In Paper I we used the  Galactic extinction model by \citet{are92}. In
this study, however, we wished to  test the sensitivity of the derived
luminosities  on  the extinction  model  by  using other  more  recent
Galactic extinction models: the Bayestar19 model based on the E$(B-V)$
colour excess  \citep{gre19}, which was  also used in the  \Gaia\ EDR3
collaboration  \citep{ant21};  the  \Gaia-2MASS 3D  Galactic  maps  by
\citet{lal19},  and  the  Galactic  model by  \citet{dri03}.  For  the
reddening corrections,  we used  the relations $A_K  = 0.114  A_V$ and
$A_J =  2.47 A_K$,  except when  we used  the \citet{gre19}  model, in
which case we  used the corrections given in  \citet{ant21} study (see
Table  B.1.  in  the  appendix).   The  3D  dust-reddening  maps  from
Bayestar19  \citep{gre19}  are  based  on \Gaia\  DR2  parallaxes  and
stellar photometry  from Pan-STARRS 1  and 2MASS. This map  covers the
sky north of a declination of $-30^{\circ}$ out to a distance of a few
kiloparsec.  The  spatial limitation  means  that  between $10\%$  and
$30\%$  of the  sample  stars  do not  fall  within  the selected  sky
coverage. The 3D absorption map from \citet{dri03} provides the V-band
absorption using scaling factors  up to $10\,$kpc. Absorptions towards
the  inner disc  are known  to be  overestimated, however.  \Gaia\ DR2
photometric  data  were combined  with  2MASS  measurements to  derive
extinction in \citet{lal19} up to $3\,$kpc.  We used \citet{lal19}
  in combination with  \citet{mar06} in the outer  disc. A correction
in  the  latter  was  applied  to  allow  a  smooth  transition.  This
combination was applied in the last version of \Gaia\ Object Generator
\citep{Luri14} and  published in  the \Gaia\ archive.   Differences in
the $A_V$ values  estimated from the \citet{are92}  Galactic model and
those from    \citet{gre19}, \citet{lal19},   and
\citet{dri03} models for the  stars  in common  with Paper  I  are not  very
significant. For the $K_{s_{o}}$ magnitude, we found a mean difference
of $-0.02\pm0.08$, $-0.03\pm0.06$, and $0.04\pm 0.10$ mag in the sense
\citet{are92} minus \citet{gre19}, \citet{lal19}  and \citet{dri03} magnitudes,
respectively,  while  for the  $J_o$  magnitude,  the differences  are
$-0.046\pm0.2$,    $-0.085\pm   0.15$,    and   $-0.12\pm0.23$    mag,
respectively. We do not find any correlation of these differences with
the distance of the object. We  note that most of the dispersion found
in these differences is produced by  a small number of stars for which
there is a large discrepancy in the $A_V$ value estimated according to
\citet{are92}  and the  three alternative  Galactic extinction  models
used here.

Figure 6  shows the comparison  between the $A_V$ values  derived from
the  \citet{lal19}   model  and  those  from   the  \citet{gre19}  and
\citet{dri03}  models  for  the  full  sample  of stars  in  this
  study.  The  extinction  values   derived  from  \citet{lal19}  and
\citet{dri03} agree  fairly well for  $A_V\lesssim 1.7$ mag;  the mean
differences are  $0.00\pm 0.40$ mag  in the sense  \citet{lal19} minus
\citet{dri03}. For higher values,  $A_V\gtrsim 1.7$ mag, there is
  a  significant discrepancy  between  the two  models, although  this
  affects  a limited  number of  objects in  our sample.  Sources with
  $A_V\gtrsim 1.7$  mag have heliocentric distances  larger than $\sim
  2-3$ kpc, precisely where  \citet{mar06} overtakes the new estimates
  by  \citet{lal19} and,  as shown  by \citet{mar06},  the differences
  with   \citet{dri03}  are   as   expected.   The  comparison   with
\citet{gre19} (bottom  panel in  Fig. 6) reveals  the same  pattern: a
mean  difference of  $-0.20\pm 0.40$  mag for  $A_V\lesssim 1.7$  mag,
while for larger $A_V$  , \citet{lal19} obtained systematically higher
values, but  again this affects a  limited number of objects.  We note
that  the  overwhelming  majority  of  the  objects  for  which  large
differences  in $A_V$  are  found  are located  at  very low  Galactic
latitude and  at heliocentric  distances larger than  $2.5\,$kpc. This
distance approximately coincides with the transition to the outer disc
where  \citet{lal19}   combines  with  the   \citet{mar06}  extinction
values. In any  case, although the choice of the  extinction model may
produce significant differences in  the luminosity estimation for some
particular objects, we confirmed  that the global luminosity functions
(see  below) are  not  significantly affected. The  luminosity
  function  is  obtained  from  the $K_s$  magnitude,  which  is  only
  slightly  affected by  Galactic  extinction. For  this reason,  we
finally  decided to  use  the interstellar  extinctions obtained  from
\citet{lal19}  (in combination with \citet{mar06})  that allowed
an estimate  of $A_V$ for  $100\%$ of the  stars in the  final sample,
while $\sim  95\%$ and $\sim 70\%$  can be determined in  the cases of
the        \citet{dri03}        and       \citet{gre19}        models,
respectively. Unfortunately, the \citet{lal19} model does not estimate
the uncertainty in $A_V$ for individual objects.

\begin{figure}
\centering
\includegraphics[width=9cm,angle=-00]{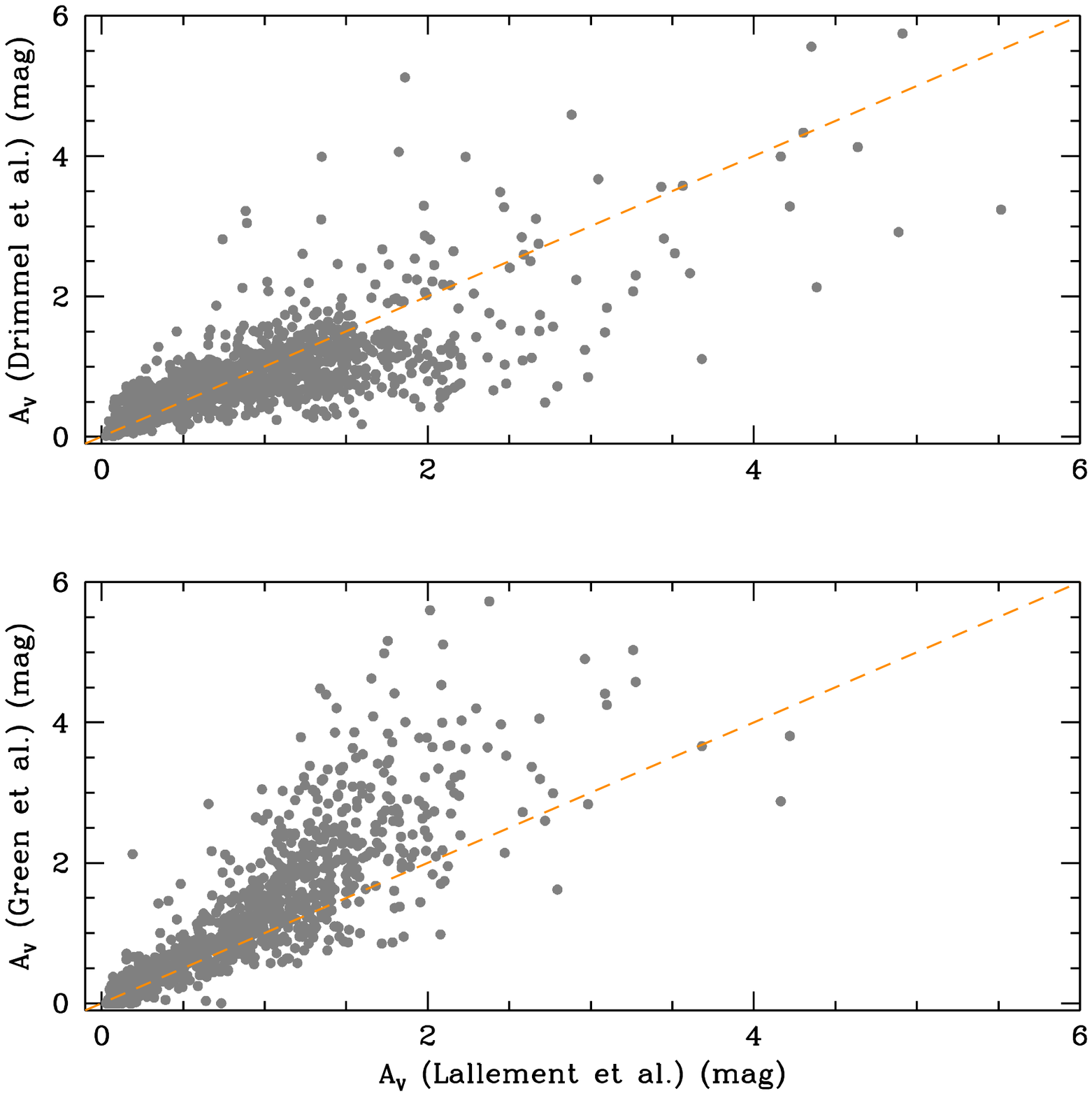}
\caption{Comparison of the interstellar extinction $A_V$ derived from the different Galactic extinction
  models used in this study. The dashed orange line shows the 1:1 relation.}
\end{figure}

The corrected  $J$ and $K_s$ magnitudes  are given in Table  1 for all
the sample stars. M$_{K_{s}}$  and M$_{\rm{bol}}$ magnitudes were then
derived  from   the  distance   modulus  relation.   Uncertainties  in
M$_{K_{s}}$  and   M$_{\rm{bol}}$  are  dominated  by   those  in  the
distances. For  the typical parallax  uncertainty in our  stars ($\leq
10\%$, see  Sect. 2), an error  of $\sim\pm 0.20$ mag  in the absolute
magnitudes is estimated. Adding  the uncertainties associated with the
$J$ and $K_s$  magnitudes and the bolometric  correction, we estimated
$\pm 0.25$ mag  as a typical error for  M$_{K_{s}}$ and M$_{\rm{bol}}$
in  the full  sample  of  stars. The  actual  uncertainty is  probably
slightly higher because  we did not consider the  uncertainty in $A_V$
in this estimate (see  above). The corresponding bolometric luminosity
distributions and luminosity functions (LF) obtained for all the giant
carbon star types in our sample are shown in Figure 7. We note that to
construct the  final LF for  normal N-type stars (blue  histogram, top
panel in  Fig. 7), we  excluded the stars  that were classified  as of
unknown type (36 objects) by \citet{che12}, even though they are
quoted as possible carbon stars  in the SIMBAD database. Indeed, $\sim
30\%$ of these objects are  classified as RCrB type stars,
and another  $\sim 10\%$  as binary  or multiple
  stellar systems in  SIMBAD. The  R  Coronae   Borealis  (RCrB)  stars  are   rare  hydrogen-deficient
  carbon-rich supergiants  that are  best known for  their spectacular
  declines  in brightness  at  irregular  intervals. Two  evolutionary
  scenarios  have   been  suggested  to   produce  an  RCrB   star:  a
  double-degenerate  merger of  two white  dwarfs, or  a final  helium
  shell  flash   in  a   planetary  nebula  central   star  \citep[see
    e.g.][]{cla12}.  Furthermore, we also  excluded the extreme
red  objects ($(J-K_s)\geq  2.5$ mag,  191 objects)  in the  sample of
\citet{che12}  because  BC$_K$ is  uncertain  for  these stars.  These
stars, moreover, may have relatively high mass-loss rates (see below),
and  the  resulting circumstellar  extinction  causes  them to  appear
fainter in  the $K_s$ band. We  thus derived the luminosity  of 300 of
the initial 491 candidate normal AGB carbon stars.  From Fig. 7 we see that:\\

\begin{figure}
\centering
\includegraphics[width=9cm,angle=-00]{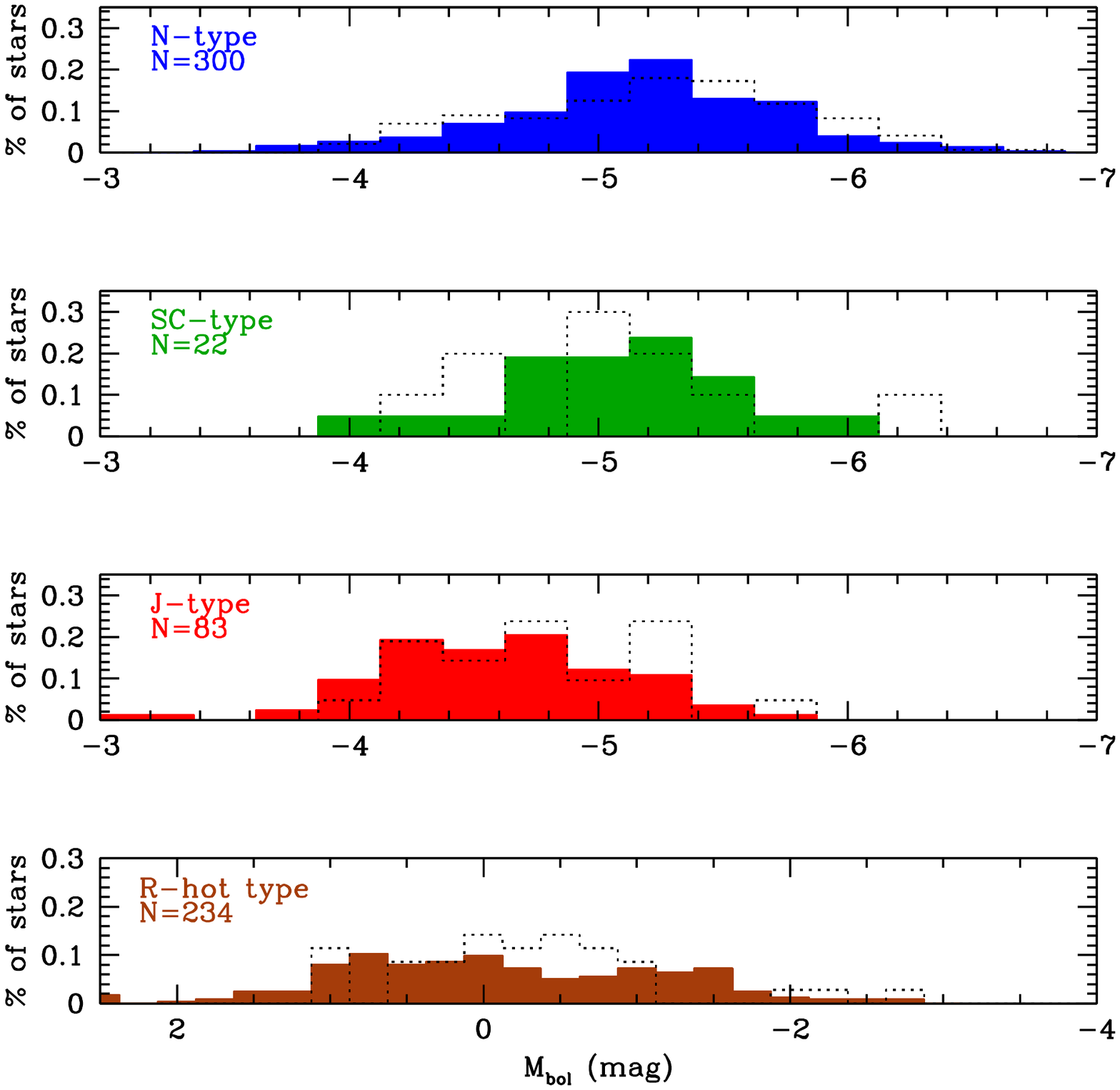}
\caption{Luminosity distributions (coloured histograms) derived in this study for the different
  spectral types of Galactic carbon stars (CH-type stars are not shown). The bin size is 0.25 mag.
  The range in luminosity is different for the R-hot stars that are fainter. For comparison, the corresponding
histogram obtained in Paper I for each spectral type is shown in each panel (dotted histograms).}
\end{figure}

a)     The    average     luminosity     of     N-type    stars     is
$\langle$M$\rm{_{bol}}\rangle= -5.04\pm 0.55$ mag. This almost matches
the average  value obtained  in Paper  I. For  comparison, we  show in
Fig. 7  (dotted lines) the  corresponding LF  obtained in Paper  I for
each spectral  type. For N-type stars,  the new LF appears  to be more
symmetric around the average value: a Kolmogorov-Smirnov test confirms
that the distribution is consistent  with a Gaussian distribution with
a $p=0.75$. It  also shows much less relevant tails  at low and mainly
at high  luminosities. We discuss  the significance of these  LF tails
below. On the  other hand, the derived average  absolute $K$ magnitude
is $\langle$M$_{K_s}\rangle=-8.16\pm 0.57$ mag,  which is identical to
the  value derived  in Paper  I and  agrees with  the values  found in
near-IR    photographic    surveys    of   the    Magellanic    Clouds
\citep[e.g.][]{fro80} and  the Galaxy  \citep{sch87}. As  commented in
Paper I, the dispersion in  M$_{K{_s}}$ is compatible with the typical
range  in  T$_{\rm{eff}}$  (2500-3500  K)  deduced  for  N-type  stars
\citep[e.g.][]{berg01}.

In Paper I we compared (see Fig. 5 and Sect. 3.1 there) the derived LF
for N-type stars  with the most recent theoretical  predictions of the
C-rich phase  on the AGB  \citep[see e.g.][]{cri15}. A  good agreement
was found between the observed  and predicted LF. However, we remarked
then  that  the existence  of  two  extended  tails  at high  and  low
luminosity would  contradict standard modelling  of the AGB  phase, in
particular, with the lower and upper limit of mass to which a star can
become C-rich during the AGB phase  ($\sim 1.5$ M$_\odot$ and $\sim 3$
M$_\odot$, respectively).  The new  LF derived here  (see Fig.  4, top
panel),  based on  a better  statistics and  more accurate  distances,
shows less extended  and significant tails (in  particular for highest
luminosities), which  alleviates this  contradiction. In any  case, we
demonstrated   in  Paper   I  that   the  low-luminosity   tail  might
theoretically be accounted for if a small fraction of the N-type stars
were  extrinsic   (binary)  stars   \citep[an  idea   first  suggested
  by][]{iza04}, while the high-luminosity tail would be more difficult
to  reproduce.  \citet{sta05}  showed  that  using  a  high  mass-loss
prescription during the AGB, the carbon  star phase can be more easily
attained   for  the   most  massive   models,  those   populating  the
high-luminosity tails of the  LF. Nevertheless, the observed existence
of N-type stars with M$\rm{_{bol}}\lesssim -5.5$ mag, which we confirm
here, severely  constrains the  occurrence of  the hot  bottom burning
(HBB)    in    intermediate-mass    AGB    stars. Although
  intermediate-mass models show that the  HBB ceases when the envelope
  mass is reduced to $\sim 1$ M$_\odot$ \citep[see e.g.][]{kar14}, the
  duration  of the  C-star phase  would be  too short  to provide  any
  sizeable  contribution to  the LF,  but very  bright C-stars  indeed
  exist    in   the    LMC    with    M$_{\rm{bol}}\sim   -6.8$    mag
  \citep{loo99,tra99}.

b) The larger number of SC-type stars we analysed here with respect to
Paper I allows us to confirm that their LF function is very similar to
that  of  the  N-type  stars.  Moreover,  the  very  similar  chemical
composition    shared    by    N-    and    SC-type    carbon    stars
\citep{abi98,abi02,abi19}, as well as their similar location above and
below the Galactic plane (see Section 2), supports the conclusion that
both types of carbon stars originate from similar progenitors. This is
compatible with  the hypothesis  that the  SC-type represents  a short
transition phase (C/O $\approx 1$) in between the O-rich (C/O$<1$) and
the  C-rich (C/O$>1$)  AGB phase.  The average  luminosity values  are
$\langle$M$\rm{_{bol}}\rangle=     -4.90\pm      0.46$     mag     and
$\langle$M$_{K_s}\rangle=-8.00\pm 0.44$ mag.

c) The new  LF for J-type stars  is about $0.2$ mag  dimmer on average
compared to that derived in Paper I. We ascribe this difference to the
larger statistic  used here. Now  it is clear  that the LF  for J-type
stars  is dimmer  than  that  of N-type  carbon  stars  by $\sim  0.5$
mag.          Their          average         luminosities          are
$\langle$M$_{\rm{bol}}\rangle=-4.42\pm       0.53$      mag,       and
$\langle$M$_{K_s}\rangle=-7.53 \pm 0.57$  mag. However, although these
luminosites  are within  the range  expected during  the AGB  phase, a
non-negligible  fraction   of  the   J-type  stars  is   fainter  than
M$\rm{_{bol}}\sim -4.5$  mag, which  represents approximately the threshold  for the
occurrence         of  third dredge-up  (TDU)         episodes         \citep[see
  e.g.][]{stra03b,cri11}.  Therefore,  we  confirm our  conclusion  of
Paper I that the origin of  J-type stars (as their carbon enhancement)
is different from that of N-type stars.

\begin{figure}
\centering
\includegraphics[width=9.5cm,angle=-00]{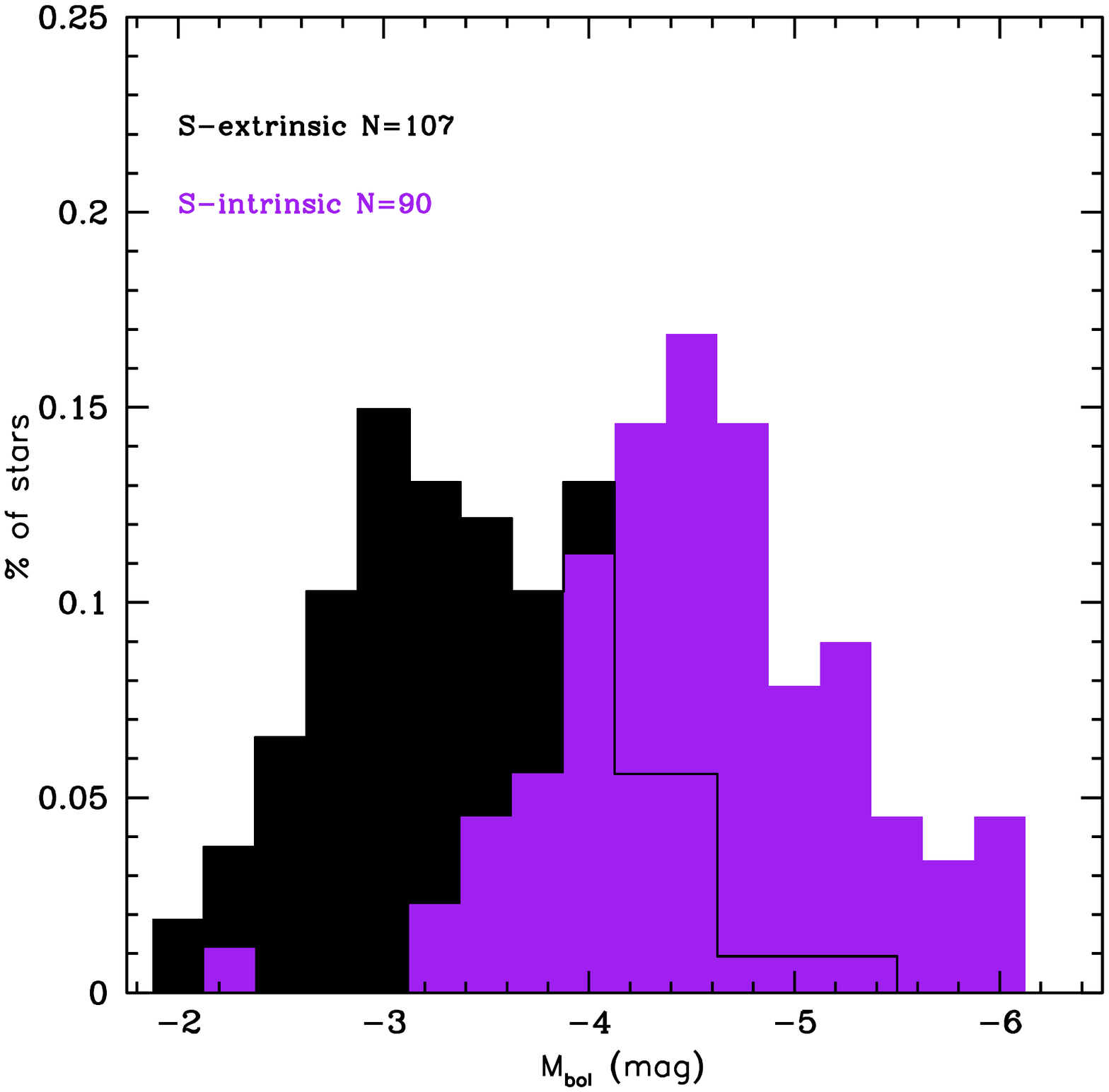}
\caption{Luminosity distributions derived for the S stars of this study. Purple shows intrinsic
  S stars, and black shows extrinsic S stars. The bin at  M$_{\rm{bol}}\sim -2.0$ mag.
  corresponds to the star Hen 4-16 alone, which is quoted in SIMBAD as a possible
  S star. The bin size is 0.25 mag.}
\end{figure}

d)   For   R-hot   carbon   stars,   we   find   average   values   of
$\langle$M$_{\rm{bol}}\rangle=0.05\pm       1.10$        mag       and
$\langle$M$_{K_s}\rangle=-2.47\pm 1.20$ mag,  which is slightly dimmer
than   those  found   in  Paper   I,  but   agrees  well   within  the
uncertainties.  This clearly  discards  the suggestion  that the  bulk
population of  these objects  consists of  He-burning red  clump stars
\citep{kna01}  because  the luminosity  of  the  red clump  for  solar
metallicity stars is  expected at $\langle$M$_{K_s}\rangle\sim -1.6\pm
0.3$ mag \citep[e.g.][]{alv00,cas00,sal02}.   The derived luminosities
place  the R-hot  stars throughout  the RGB  phase. When  we assume  a
typical mass  of $\sim  1.0$ M$_\odot$  and Z$\sim$  Z$_\odot$ stellar
models,  the bulk  of  the  observed R-hot  stars  would instead  have
luminosities that are typical of the  RGB bump and would vary from the
luminosity of the first dredge-up almost  to the luminosity of the RGB
tip.  This  wide  luminosity  range is  difficult  to  associate  with
internal processes of  mixing that are able to enrich  the envelope in
carbon  during the  RGB.  Therefore, the  hypothesis  of an  extrinsic
origin (or triggered  by a merger) of the carbon  enhancement in these
stars seems to be reinforced. We  note, nevertheless, that there is no
observational evidence of binarity in these stars \citep{mac84}.

\begin{figure}
   \centering
   \includegraphics[width=9.5cm]{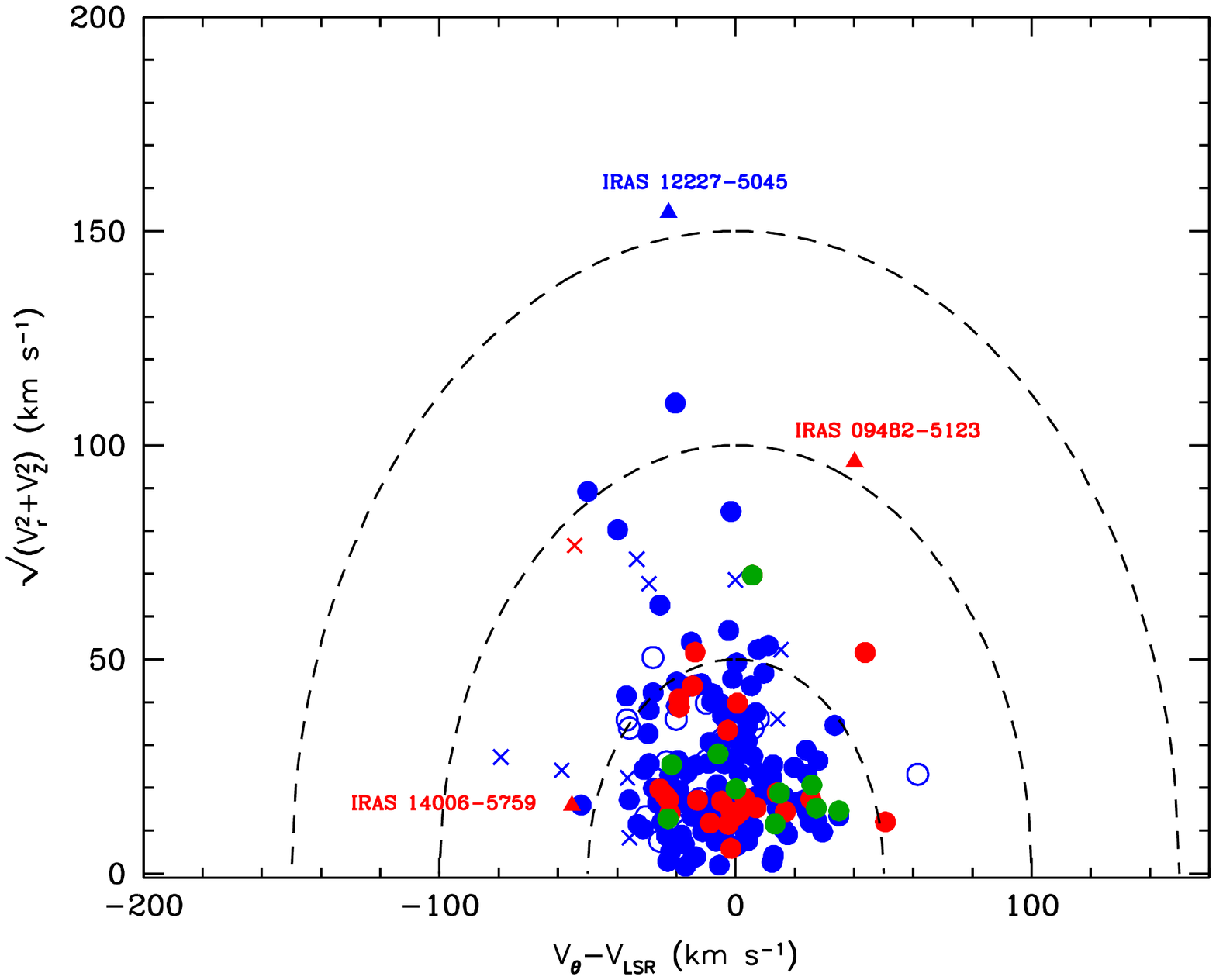}
   \caption{Toomre diagram for the selected carbon stars of this study. The colour
     symbols are the same as in Fig. 1. Stars are plotted according to their membership
     probability ($>80\%$) to belong to the thin disc (solid circles), thick disc
     (crosses), or halo (triangles) stellar population. Open circles show stars with
     ambiguous membership (see text). Dashed lines indicate $V_{\rm{tot}}=\sqrt{ V_r^2+V_Z^2+(V_\theta-V_{\rm{LSR}})^2} = 50, 100,$
     and 150 km s$^{-1}$. Some stars with kinematics compatible with the Galactic halo are labelled (see text). }
  \end{figure}

Finally, Figure 8  shows the LF derived for the  extrinsic (black) and
intrinsic  (purple) S  stars  from the  \citet{che19}  survey. To  our
knowledge, this LF  is derived in such  a large number of  S stars for
the     first      time.     The     average      luminosities     are
$\langle$M$_{\rm{bol}}\rangle=-4.42\pm            0.68$           mag,
$\langle$M$_{K_s}\rangle=-7.50\pm        0.79$         mag,        and
$\langle$M$_{\rm{bol}}\rangle=-3.52\pm            0.68$           mag,
$\langle$M$_{K_s}\rangle=-6.40\pm   0.75$   mag  for   intrinsic   and
extrinsic S stars, respectively. It  is evident that intrinsic S stars
have higher luminosities  than the extrinsic ones  and, in particular,
the  luminosities of  the overwhelming  majority are  higher than  the
predicted  onset of  the TDU  during  the AGB.  For solar  metallicity
stars,  this luminosity  limit is  M$_{\rm{bol}}\sim -3.5$  mag, which
corresponds  to  stars  with  an  initial  mass  above  1.3  M$_\odot$
\citep[see e.g.][]{sie00,cri11,kar14,esc17}. This result is consistent
with intrinsic (Tc-rich) S stars being TP-AGB (thermal pulsing) stars, which agrees with
the  conclusion recently  reached  by \citet{she21},  who derived  the
luminosities of S  stars from theoretical fits to  the spectral energy
distributions.  Moreover,  we  also  found a  considerable  number  of
high-luminosity (M$_{\rm{bol}}\leq  -5.5$ mag) intrinsic S  stars (see
Fig. 8).  This high luminosity  is compatible with these  stars having
M$> 3-4$  M$_\odot$. Stars with such  a high mass may  experience HBB,
which prevents them from becoming C-rich AGB stars. On the other hand,
the  derived LF  of extrinsic  S stars  (peaking at  M$_{\rm{bol}}\sim
-3.0$  mag) is  compatible  with these  objects  (binaries) being  1-2
M$_\odot$   stars  in   the   RGB   phase  \citep{van98,she18}.    The
high-luminosity  (M$_{\rm{bol}}\leq  -4.0$  mag)  tail in  the  LF  of
extrinsic S  stars (see Fig.  8) may  correspond to barium  stars with
masses in excess of $\sim 2.5$  M$_\odot$ that can turn into extrinsic
S  stars  on  the  early  AGB,  but  they  are  expected  to  be  rare
\citep[see][]{she18}.

\section{Kinematics}

\begin{figure}
   \centering
   \includegraphics[width=9.5cm]{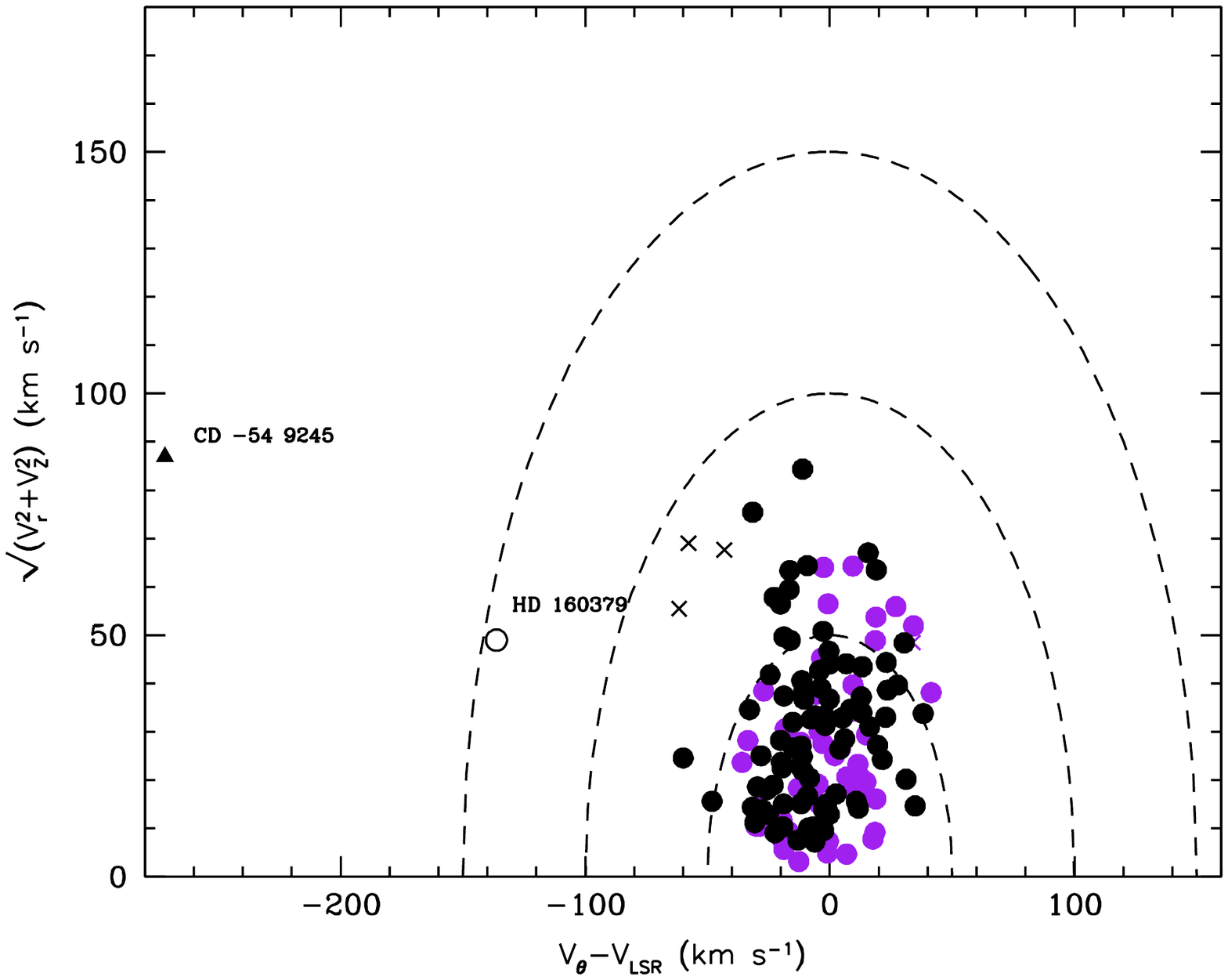}
   \caption{Same as Fig. 9 for extrinsic (black) and intrinsic (purple) S stars. The star HD 160379
   (open circle) has an ambiguous Galactic membership population between thick disc and halo according to our criteria (see text).}
  \end{figure}

\Gaia\ EDR3 astrometric data provide  an additional and valuable piece
of information  to fully  characterise the  stellar population  of the
different  carbon star  types. As  done in  Paper I,  we computed  the
galactocentric positions and velocities of  our sample stars using the
line-of-sight distance  estimates of  \cite{bai21}, together  with the
RA,  DEC,  and   proper  motions  of  \Gaia\  EDR3   and  our  adopted
line-of-sight velocities  (V$_{\rm{rad}}$, see Sect.  2). Furthermore,
we assumed  the following  parameters: (R$_\odot$, Z$_\odot)  = (8.34,
0.025)$  kpc \citep{rei14},  (U$_\odot$,  V$_\odot$,  and W$_\odot)  =
(11.1,  12.24,  \text{and  } 7.25)$  km\,s$^{-1}$  \citep{sch12},  and
V$_{LSR}=240$ km\,s$^{-1}$ \citep{rei14}.
For details  of the  estimation of  the uncertainties  on each  of the
derived parameters, we  refer to Paper I. The  final computed velocity
components, Z  coordinate, and  galactocentric positions of  the stars
are reported in Table 1.

 Figures 9,  10, and  11 show  the Toomre  diagrams for  the different
 spectral  type stars  in this  study. For  clarity, we  separated the
 sample stars  into three  different plots,  corresponding to  the AGB
 carbon stars (Fig. 9), S-type stars  (Fig. 10), and CH-type and R-hot
 stars (Fig. 11). We excluded from  these diagrams the stars listed as
 possible carbon  stars in \citet{che12}, and  of possible
 N-type and  as unknown  carbon type in  \citet{ji16}. Following
 the method outlined in Paper I,  we computed the likelihood to belong
 to each of the Galactic components  for each star. If this likelihood
 was greater  than 80\%,  we assigned the  membership of  the $i^{th}$
 component that  scored this probability  for this star.  According to
 this  criterion, stars  are listed  with a  0, 1,  and 2  in Table  1
 depending on whether they belong to  the thin or thick disc and halo,
 respectively. Stars  labelled with a  3 in this table  have ambiguous
 membership according  to our criterion. %In  Figs. 6, 7 and  8, stars
 belonging to the thin disc are plotted with solid circles; thick disc
 with crosses; halo  with solid triangles, and  open circles represent
 stars with ambiguous membership.

We confirm the figure found in Paper  I in the sense that N-, SC-, and
J-type carbon stars mostly belong to the thin disc, with a probability
rate of 97\%,  100\% and 90\%, respectively.  A few  N-type stars have
typical thick-disc  kinematics, and the star  IRAS 12227-5045 probably
belongs  to the  halo.  This star,  together  with V  CrB  and RU  Vir
identified in Paper I, probably is one of the very few halo AGB carbon
stars  known   in  our   the  Galaxy.  They   deserve  high-resolution
spectroscopic    studies   to    define   their    detailed   chemical
composition. The same remark may be applied to two J-type stars with a
probable  halo membership:  \object{IRAS 14006-5759}  and \object{IRAS
  09482-5123} (see Fig. 9).

On the other  hand, Fig. 10 also shows that  the overwhelming majority
of extrinsic  and intrinsic S  stars belong to the  Galactic thin-disc
population (probability  rates of 94$\%$ and  96$\%$, respectively). A
few extrinsic  S stars belong to  the thick disc, and  we identify the
star \object{CD -54  9245} as a possible halo star.  We point out that
\citet{vant07} did not detect Tc or Li lines in this S star.

Finally, Figure  11 shows  the Toomre  diagram for  the CH-  and R-hot
types (the  latter type also  includes those  studied in Paper  I). As
expected, a  significant fraction ($\sim  56\%$) of the  CH-type stars
belong to  the halo  and/or thick-disc  population, but  a significant
number ($\sim  44\%$) also belongs  to the  thin disc.  About  15\% of
the\  R-hot  stars  belong  to  the   thick  disc,  and  two  of  them
(\object{HIP   40374}    and   \object{J230231.75+210129.7})   exhibit
kinematics that are fully compatible  with the halo. The kinematics of
at least 20 additional R-hot stars (open brown circles in Fig. 11, see
also Table 1) are compatible with  the thick disc, and the rest ($\sim
50\%)$  exhibit  thin-disc  kinematics.  Because  the  V$_z$  velocity
component increases  with the stellar age  \citep[e.g.][]{nor04}, this
figure reinforces our conclusion of Paper I: a significant fraction of
the R-hot  stars belongs to  an older (probably less  massive) stellar
population than  the other types of  giant carbon stars. Fig.  11 also
clearly shows that  CH- and R-hot types share  rather similar Galactic
kinematics. In  the next  section, we  show that  they also  have very
similar  luminosities (M$_{K_s}$).  The  conclusion  is tempting  that
there is a link between these two types of carbon-rich stars. However,
most  of  the CH-type  stars  are  metal-poor binary  objects  showing
s-elements enhancements in their surface, but none of these properties
are observed in R-hot stars \citep{zam09}.

\begin{figure}
   \centering \includegraphics[width=9cm]{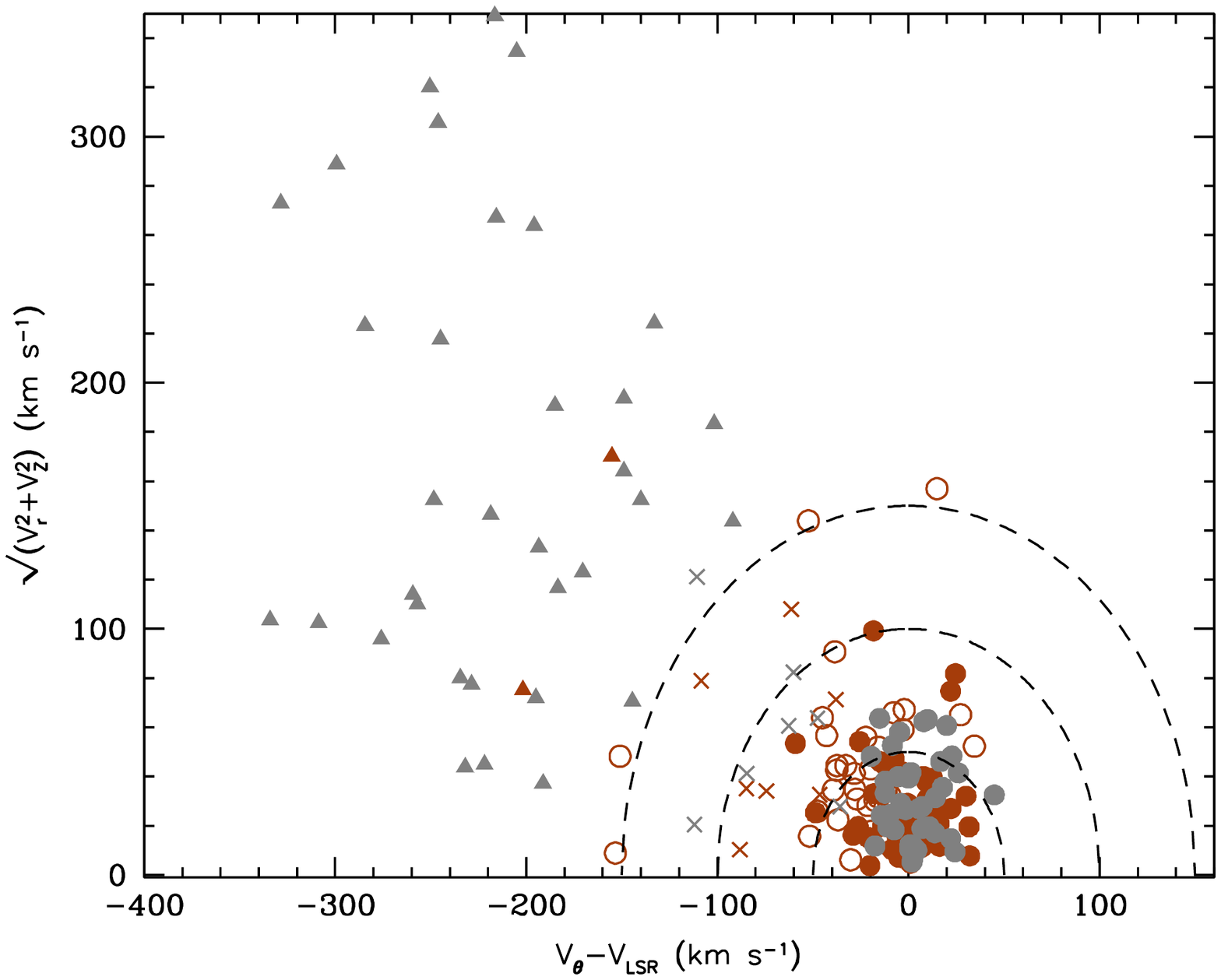}
   \caption{Same as Fig. 9 for the  CH-type stars in the LAMOST sample
     \citep{ji16} together  with the  R-hot stars  also in  LAMOST and
     those   included   in   Paper   I  (grey   and   brown   symbols,
     respectively). A significant number of stars classified in LAMOST
     as R-hot type have ambiguous Galactic population membership (open
     brown  circles), mainly  between  the thin  and  thick disc  (see
     text).}
  \end{figure}
  
  \begin{figure}
   \centering \includegraphics[width=9cm]{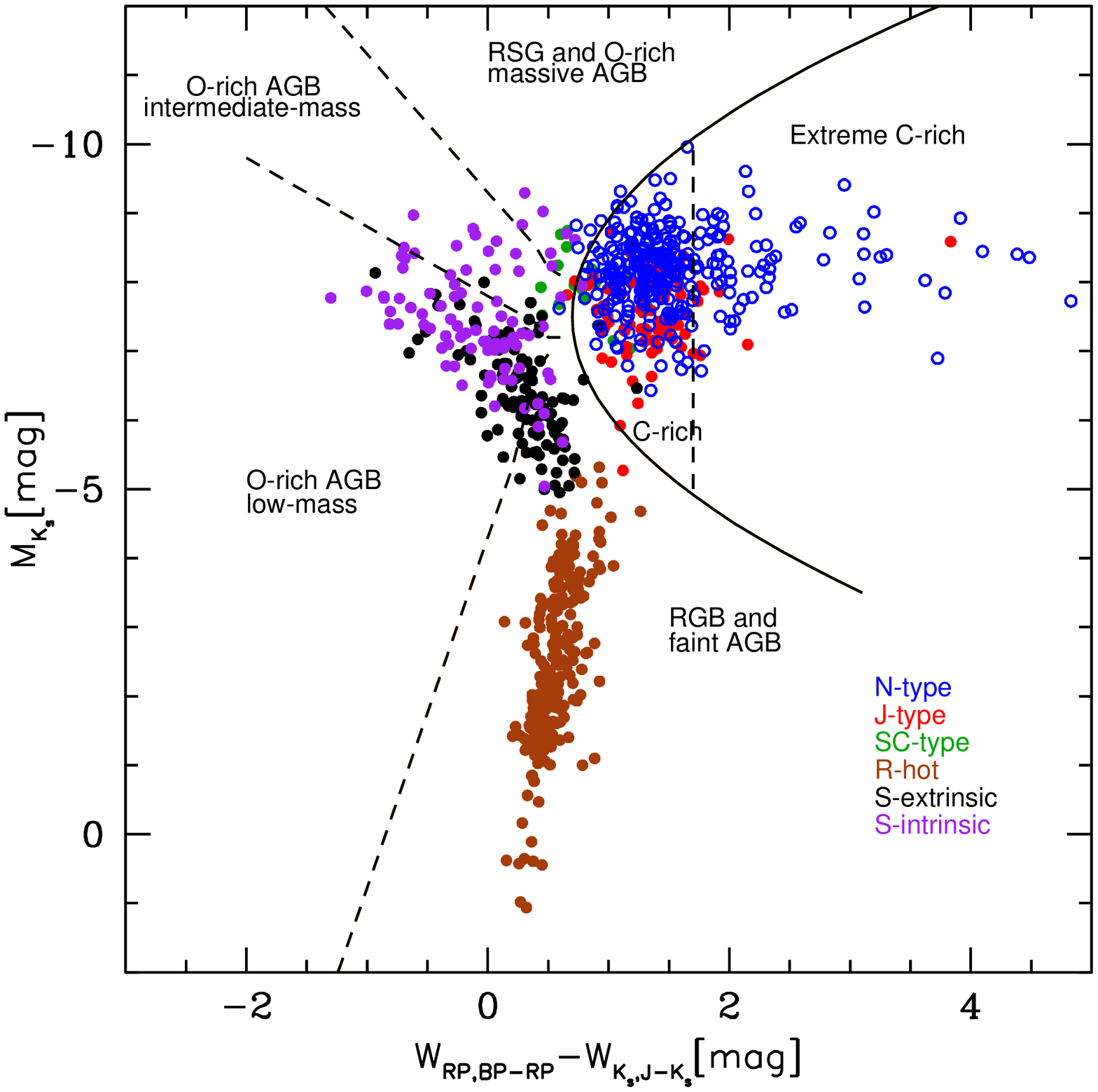}
   \caption{\Gaia-2MASS diagram  for our main star  sample. The curved
     line delineates the theoretical limit between O-rich (left of the
     line)  and C-rich  AGB stars  (right of  the line).  Dashed lines
     separate sub-groups  of stars  as indicated  in the  figure.  The
     colour code  for the different AGB  types is shown in  the bottom
     right   corner.  Open   blue  circles   show  N-type   stars  for
     clarity. Several  SC-type stars (green circles)  and J-type stars
     (red  circles) are  placed  in  the C-rich  zone,  but cannot  be
     distinguished in the figure  because of crowding. The uncertainty
     in M$_{K_s}$ typically is $\pm 0.23$ mag. }
   \label{Fig:Gaia-2MASS}
  \end{figure}

 \section{Identification of the stellar types in the \Gaia-2MASS diagram} 
  
 After we calculated the luminosities of the stars, we constructed the
 \Gaia-2MASS  diagram  (see Figs.  12  and  13). This  was  originally
 designed by \citet{leb18} and is especially suitable to highlight the
 presence of AGB stars.  In this diagram, the M$_{K_{s}}$ magnitude is
 correlated  with  a  particular   combination  of  \Gaia\  and  2MASS
 photometry         through        the         quantity        W$_{\rm
   {RP,BP-RP}}-$W$_{K_{s},J-K_s}$,  where   W$_{\rm  {RP,BP-RP}}$  and
 W$_{K_{s},J-K_s}$    are     reddening-free    Wesenheit    functions
 \citep{sos05},  defined as  W$_{\rm {RP,BP-RP}}=G_{RP}  -1.3(G_{BP} -
 G_{RP})$ and  W$_{K_{s},J-K_s}= K_s-0.686 (J-K_s)$,  respectively. In
 Paper I  we demonstrated  that this  diagram is  a powerful  tool for
 analysing and  identifying the  different spectral  types of  the AGB
 stars as a function of chemical type and initial mass.

Figure 12 shows that the C- and O-rich AGB stars in our study populate
different regions, which allows us to  easily identify them as long as
their distances are  known. This diagram also  permits the distinction
between regions  with low-mass, intermediate-mass, and  massive O-rich
AGB stars, RGB, or faint AGB  stars as well as supergiants and extreme
C-rich AGB  stars, the specific  stellar mass  range in each  of these
groups   depending   on   the  stellar   metallicity   \citep[see][for
  details]{gir05,mar17,leb18}. The region  called extreme C-rich stars
concerns  very  red objects  with  high  $(J-K_{s})$ values  that  are
associated  with  high mass-loss  rates.  As  shown  in Paper  I,  the
\Gaia-2MASS diagram  clearly identifies the different  types of carbon
stars. Now  this fact  becomes more apparent  because we  studied more
stars  for each  spectral  type.  We  basically  confirm our  previous
findings: N-type  stars (open blue  circles) clearly occupy  the whole
C-rich  region, while  J-type  stars  (red) are  located  in the  same
region,  but are  shifted to  fainter M$_{K_s}$,  as shown  in Section
3. Most of the SC-type stars (green), which are characterised by a C/O
ratio very close  to unity, are clearly located at  the border between
O-rich and C-rich  objects (several of them cannot be  seen in Fig. 12
because of the  crowding with N-type stars). A  considerable number of
N-type  stars  are  located  in   the  extreme  C-rich  zone,  W$_{\rm
  RP,BP-RP}-$W$_{\rm  K_s,J-K_s}>1.7$  mag.  These  stars  would  have
higher mass-loss rates \citep{mar21}.  Furthermore, Fig. 12 also shows
that several N- and SC-type stars are located in the region of massive
AGB stars (or RSG) and very  close to the C-rich region. Although they
are not very  luminous (M$_{K_{s}}\sim -9.0$ mag),  these stars should
be massive enough to undergo HBB  because this region is thought to be
occupied by  stars with  initial mass higher  than $\sim  4$ M$_\odot$
(see the discussion in \citet{leb18}). We  note that there is no clear
observational confirmation of the  chemical anomalies predicted by the
operation of the HBB in AGB stars as yet \citep{kar14, abi17}. This is
in part  due to the  observational difficulty of  identifying suitable
objects. We show that the \Gaia-2MASS diagram may be a useful tool for
this, and in  consequence, to place limit to the  minimum stellar mass
for the operation of the HBB.   On the other hand, the region occupied
by R-hot stars  (brown) is clearly defined in  the \Gaia-2MASS diagram
(the  RGB and/or  faint  AGB  area), covering  a  very  wide range  in
luminosity,  but  does   not  reach  the  RGB   tip  luminosity  limit
(M$_{K_{s}}\sim  -7.0$  mag), and  very  small  dispersion in  W$_{\rm RP,BP-RP}-$W$_{\rm K_s,J-K_s}$.

Extrinsic (black)  and intrinsic  (purple) S  stars are,  as expected,
located  in  the O-rich  regions  (see  Fig.  12).  We find  only  one
exception to  this for the  extrinsic S star  \object{TYC 3717-706-1},
which is clearly located in  the C-rich region. This star is a
  RSG candidate  \citep{mes19}. These  regions are populated  by (i)
low-mass  ($\lessapprox 1.3$  M$_\odot$) TP-AGB  stars and  faint AGB,
which do  no become carbon stars  (extrinsic stars), and (ii)  by more
massive stars  ($\gtrapprox 1.5$ M$_\odot$)  that spend part  of their
AGB  evolution  as  O  rich,  and  later  become  carbon  stars  as  a
consequence of a few TDU events  and move quickly to the C-rich region
as soon the C/O ratio approaches unity (intrinsic stars). The brighter
objects (M$_{K_{s}}<-8.5$  mag) among  the intrinsic S  stars probably
are more massive  stars ($\geq 4$ M$_\odot$) that  never become carbon
stars because of the operation of  the HBB.  Several intrinsic S stars
shown in Fig.  12 are indeed located in the  O-rich massive AGB region
and are therefore also candidate HBB stars. This difference in mass is
reflected  in  the  brighter  average  M$_{K_{s}}$  magnitude  of  the
intrinsic S stars, as Fig. 12 shows (see also Section 3).

  \begin{figure}
   \centering
   \includegraphics[width=9cm]{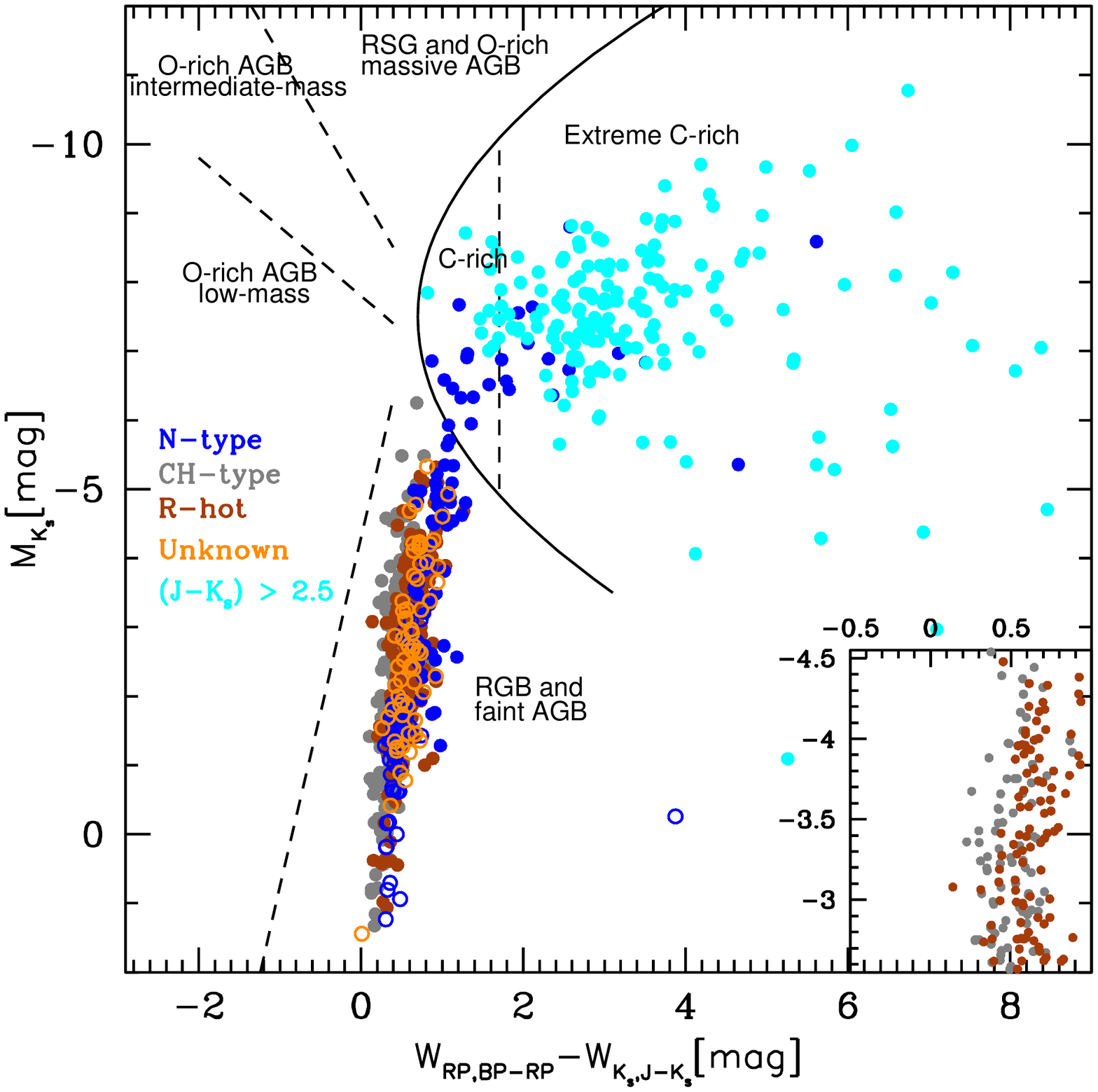}
   \caption{Same as Fig. 12 for the carbon stars in the LAMOST sample by \citet{ji16}. Grey circles
     show CH-type stars, brown circles show R-hot stars, blue circles show normal N-type stars, open
     blue circles show  possible N-type stars, and open orange circles show carbon stars of  unknown type.
     For completeness, we show (cyan circles)  the carbon stars in \citet{che12} with $(J-K_s)>2.5$ mag.
   The inset shows a zoom of a small area in the location of the R-hot and CH-type stars (see text).}
  \end{figure}
  
 Figure 13 shows the corresponding  \Gaia-2MASS diagram for the carbon
 stars identified  in the LAMOST  sample \citep{ji16}. We  recall that
 the spectral type classification in this  study was made on the basis
 of several spectral line  indices using low-resolution spectra. These
 authors  identified carbon  stars of  N,  R-hot, CH  type, and  other
 objects classified by  them as possible N type and  of unknown carbon
 type. These objects  are plotted in Fig. 13 as  blue, brown, and grey
 circles or blue and orange open circles, respectively.  Fig. 13 shows
 that the overwhelming  majority of the stars classified as  N type in
 the  LAMOST sample  are placed  in the  region of  RGB and  faint AGB
 stars. They are therefore very probably  not TP-AGB stars. Only a few
 of them with  M$_{K_{s}}\lesssim -7$ mag might  be TP-AGB carbon-rich
 stars.  Those located in the C-rich region that are fainter than this
 value could be  J-type stars (see Fig. 12).  A detailed spectroscopic
 analysis  would confirm  or discard  this.  The rest  of these  stars
 located in the RGB and faint AGB region are instead probably R-hot or
 CH-type stars. Real  R-hot and CH-type stars share  the same location
 in the  \Gaia-2MASS diagram because  both spectral types show  a very
 similar  range  of  luminosities  and  W$_{RP,BP-RP}-$W$_{K_s,J-K_s}$
 values. Interestingly, a zoom of the  region occupied by R-hot and CH
 stars  (see the  inset in  Fig. 10)  reveals a  small shift  to lower
 (bluer) W$_{RP,BP-RP}-$W$_{K_s,J-K_s}$  values for the  CH-type stars
 with respect  to the R-hot  stars: an  average of $0.42\pm  0.14$ and
 $0.55\pm0.17$ mag,  respectively. Considering that most  of the R-hot
 stars have near solar  metallicity \citep{zam09}, this difference may
 be due to  the typically lower metallicity of the  CH stars. However,
 as noted before,  even though CH- and R-hot stars  are located in the
 same region  in this diagram,  we cannot conclude that  CH-type stars
 are the metal-poor counterparts of R-hot  stars (or vice verse) as CH
 stars are binary stars and  show strong s-element enhancements. These
 characteristics are not observed in R-hot stars.  The location of the
 stars listed as   possible N-type and  unknown carbon type
 (open  blue and  orange  circles  in Fig.  13,  respectively) in  the
 \Gaia-2MASS  diagram indicates  that  these stars  are probably  also
 R-hot and/or  CH-stars. A detailed  spectroscopic study is  needed to
 distinguish their  specific nature. As mentioned  before, these stars
 were not  considered in the calculation  of the LF of  the AGB carbon
 stars (N type) of Section 3.
  
Finally,  in  Fig.  13  we  also show  (cyan  circles)  the  very  red
(($J-K_s)>2.5$ mag) IRCSs objects that were classified as N-type stars
in \citet{che12}.  Clearly, these objects  are located in  the extreme
C-rich region in the \Gaia-2MASS diagram where it is expected that the
carbon stars stars  develop a high mass-loos rate. The  fact that many
of  them  are significantly  fainter  that  the normal  N-type  stars,
showing  a  high dispersion  ($\langle$M$_{K_s}\rangle=-7.16\pm  1.60$
mag, see Section  3), would indicate that these stars  may be severely
affected  by  circumstellar  extinction,  which we  did  not  consider
here.  The circumstellar  extinction  is compatible  with these  stars
having high  mass-loss rates.  We recall that  we also  excluded these
stars in  the determination of  the LF  of N-type stars  because their
BC$_K$ is uncertain.

\section{New carbon star candidates}
  \begin{figure}[t]
   \centering
   \includegraphics[width=9cm]{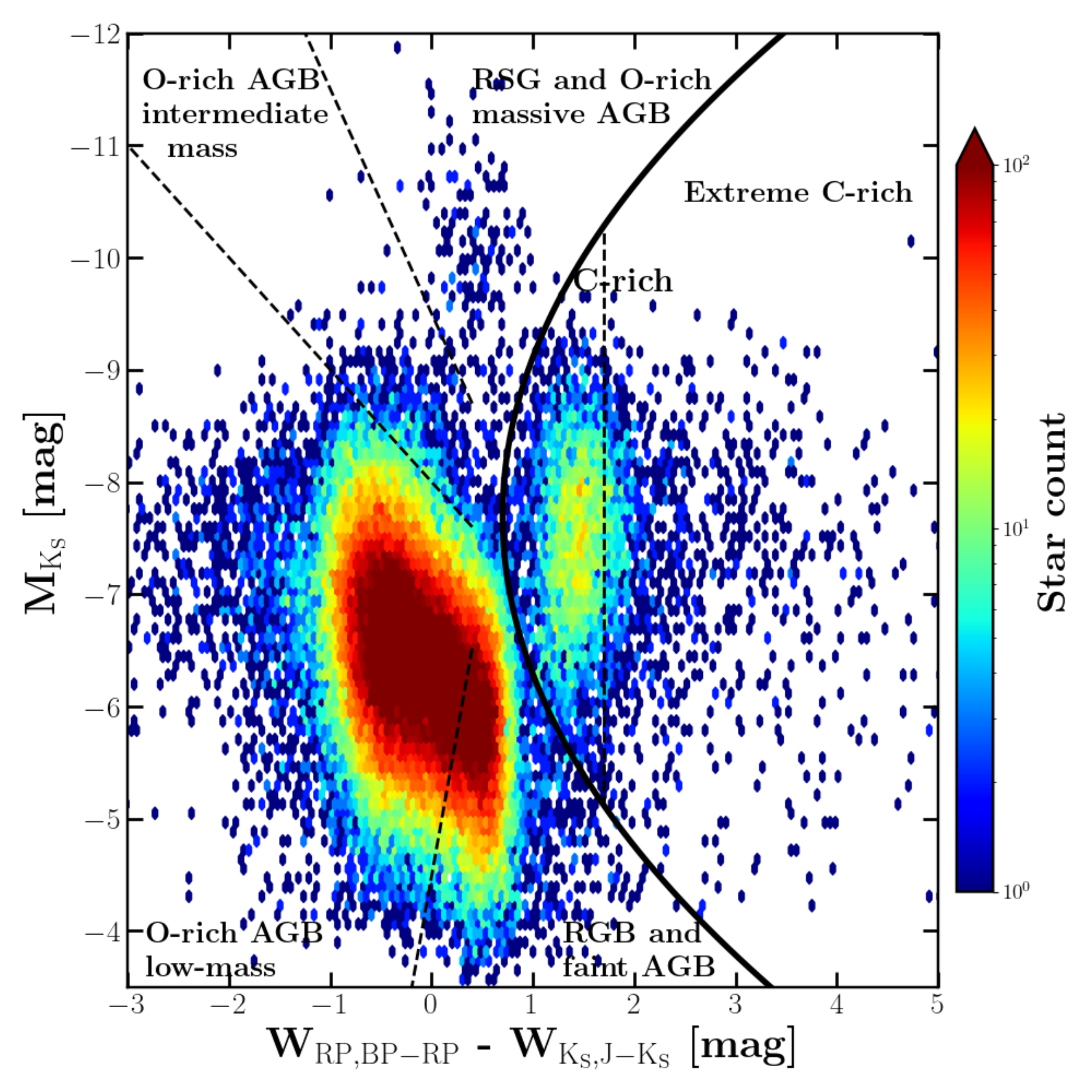}
   \caption{Same as Fig.~\ref{Fig:Gaia-2MASS}, but for new Galactic giant star
     candidates identified through their 2MASS colour and \Gaia\ parallax with
     the condition $A_{proxy}>0.06$ for the variability proxy and having M$_{K_{s}}< -4.0$ mag
     (see text).  The star count is indicated in colour code on the right side of the figure.}
   \label{Fig.NewCandidates}
  \end{figure}

We have shown in the previous section and in Fig.~\ref{Fig:Gaia-2MASS}
that carbon  stars are  found in specific  regions of  the \Gaia-2MASS
diagram,  depending on  their evolutionary  stage and/or  their carbon
enrichment. In this section, we use  a diagram like this to search for
new  Galactic   carbon  star  candidates  whose   K$_s$-band  absolute
magnitude and photometric colour can be estimated reliably.

For  this  purpose, we  searched  for  red stars  with  well-estimated
distances  and available  \Gaia\ and  2MASS photometry.   We therefore
first collected about 40 million EDR3 \Gaia\ stars with a 20\% smaller
relative  error  on  their  parallax and  a  Gaia  photometric  colour
($G_\mathrm{BP}$  - $G_\mathrm{RP})>2.0$  mag. Then,  we cross-matched
this \Gaia\  star list with  the 2MASS catalogue, leading  to $\sim$29
million  stars.  We  selected  the  reddest  stars  (i.e.  those  with
$(J-K_s)>0.5$  mag) and  those with  2MASS photometric  errors smaller
than 0.1~mag.  We then  collected their  \citet{bai21} photo-geometric
distance and created  a list of about 6.8 million  of stars with known
distances.  This  initial  sample  was  filtered  according  to  their
astronometric  fidelity  factor,  and   the  Galactic  extinction  was
estimated  as  described  in  Sect.~\ref{SecSample},  allowing  us  to
compute their  absolute magnitude  in the K$_s$  band. Because  we are
mostly interested in RGB and AGB  stars, we selected only objects with
M$_{K_{s}}<-4.0$  mag, which  reduced  the set  to  about 1.6  million
stars. Furthermore, because TP-AGB carbon stars are typically variable
stars,  we  again filtered  the  sample  with  the criterion  used  by
\citet{mow19} to  detect LPV  in {\it Gaia}  DR2 data.  This criterion
uses the uncertainty $\epsilon (\overline{F}(G))$ of the mean $G$ flux
$\overline{F}(G$)   published   for   each  source,   which   contains
information on  both the uncertainties of  the individual measurements
due to noise and on the intrinsic  scatter of the flux time series due
to stellar variability.  These authors defined a  parameter to account
for this,  $A_{proxy}(G)= \sqrt{N_{obs}}  {\epsilon (\overline{F}(G))/
  \overline{F}(G)}$, where $N_{obs}$  is the number of  good (per CCD)
observations. This information is provided by {\it Gaia} EDR3 for each
object.  We selected  as large-amplitude  LPVs red  sources that  have
$A_{proxy}(G) > 0.06$ to comply with the magnitude amplitude condition
used   in  the   DR2  catalogue   of  LPVs   (see  \citet{mow19}   for
details).  Then, we cross-matched  them with the  stellar samples
  described in Sect.~\ref{SecSample} and  removed the stars in common,
  which produced a final sample of  53,058 stars in total. We further
verified that this  final sample excluded the  potential young stellar
objects that can be contaminants to LPVs \citep[see][]{mow19}.

Because  EDR3  (and  DR3)  contain   data  collected  over  34  months
(depending on  the scanning low), this  is not enough in  principle to
detect the variable  stars with the longest periods  and classify them
correctly. Thus,  we are aware  that our bright giant  sample probably
preferably contains  stars with relatively short  periods. Because the
most luminous stars on the AGB  phase tend to have the longest periods
\citep[e.g.][]{woo00},   the   sample    might   be   biased   towards
low-luminosity AGB stars.

The final  sample of  stars is  placed in  the \Gaia-2MASS  diagram in
Fig.~\ref{Fig.NewCandidates}.  In  this  figure,   we  also  show  the
theoretical  limits of  Fig.~\ref{Fig:Gaia-2MASS}  between O-rich  and
C-rich  AGB  stars  and  between the  different  sub-groups  of  giant
stars. This  allows us  to propose a  classification for  these bright
giants.     The   large    majority    of   the    stars   shown    in
Fig.~\ref{Fig.NewCandidates} (about 67\%, 35,857 stars) are located in
the low-mass  O-rich AGB  star region.  A significant  fraction ($\sim
19\%$, 10,397  stars) are  in the  RGB and/or  faint AGB  region, 3811
objects  would be  intermediate-mass O-rich  AGB stars,  and only  334
objects would be  RSG or/and massive AGB stars. On  the other hand, we
find 2,659  new stars located in  the C-rich region, and  965 of these
lie in the  region of the extremely C-rich stars.  The probable C-rich
nature of these objects was previously unknown.

We  can compare  Fig. 14  with  the corresponding  figure obtained  by
\citet{mow19} in the Magellanic Clouds (see  their Figures 4 and 7 for
the LMC  and SMC,  respectively) and  in the Milky  Way from  DR2 data
(their Figure 10). Our conclusions here  are similar: i) There are far
fewer C-rich than  O-rich stars in the Galaxy than  in the Clouds. The
low number of carbon stars compared to O-rich stars between the Galaxy
and  the  Clouds agrees  with  the  decrease  in TDU  efficiency  with
increasing metallicity, and with a  higher O abundance in the envelope
of Galactic AGB  stars on average. On the other  hand, because some of
the stars located in  RGB and faint AGB region in Fig.  13 might be CH
and/or R-hot type stars (i.e.  C-rich objects, see previous sections),
we  could derive  a lower  limit  for the  ratio between  carbon to  M
(O-rich) stars.  It is well known  that this ratio increases  with the
decreasing (average) metallicity of the galaxy. The primary reason for
this correlation is that less C needs to be dredged-up in a metal-poor
star to enable atmospheric carbon atoms  to exceed those of oxygen. We
derive a ratio $\sim 0.05$, which is very similar to the average ratio
derived  in the  disc  of M31  \citep[see  e.g.][]{van00, ham15}.  ii)
Moreover, the distribution  of O-rich stars of  the Gaia-2MASS diagram
covers  at any  given  M$_{K_{s}}$  magnitude a  much  wider range  of
W$_{RP,BP-RP}-$W$_{K_{s},J-K_{s}}$   in  the   Galaxy   than  in   the
Clouds. The  wider distribution in  both axes  of the O-rich  zones in
Fig. 14  compared with the  corresponding feature in the  Clouds would
result from the combined effect of O-rich AGB stars turning much later
into C-rich  stars in the  Galaxy, and the  existence of a  more ample
range  of  stellar metallicities  in  the  Galactic sample.  Moreover,
Fig. 14  clearly shows the  high dispersion existing in  M$_{\it K_s}$
for  the  C-rich  objects at  a  given  W$_{RP,BP-RP}-$W$_{K_s,J-K_s}$
value. Part  of this  dispersion in M$_{K_s}$  is compatible  with the
typical range in T$_{\rm{eff}}$ (2500-3500 K) deduced for N-type stars
\citep{ber01}, and to  the circumstellar extinction (which  we did not
consider  here)  preferably for  the  objects  in the  extreme  C-rich
region. However, the  mixing of carbon stars  of different populations
probably also contributes significantly  to this dispersion in M$_{\it
  K_s}$ (and also in M$_{\rm bol}$).  Following the method outlined in
Sect. 4, we have studied the  kinematics of these stars and calculated
the membership probability to the halo  and to the thin and thick disc
of  the  $2\,659$   new  carbon  star  candidates   with  a  V$_{rad}$
measurement  according   to  EDR3  ($\sim   40  \%$  of   the  sample,
i.e.  $1\,305$  objects). Figure  15  shows  the corresponding  Toomre
diagram for these carbon stars. Of this limited sample, roughly $50\%$
belong to the thin disc (blue circles  in Fig. 15), $\sim 30\%$ to the
thick disc or halo (blue crosses and triangles, respectively), and the
rest  ($\sim 20\%$)  have  an ambiguous  membership  according to  our
membership criteria  (open blue circles).  Nevertheless,  we note that
assuming a less strict likelihood percentage to assign membership to a
population (see  Sect. 4), about $  25\%$ of the stars  with ambiguous
membership would  be thick-disc and/or halo  stars. Because thick-disc
and halo  stars are  older than  thin-disc stars  on average,  many of
these C-rich objects are very probably not intrinsic AGB carbon stars,
but extrinsic C-rich  giants: stars with masses lower  than $\sim 1.5$
M$_\odot$ that have become carbon rich  through the mass transfer in a
binary system. An alternative to this would be the possibility
  that the  minimum mass for the  formation of an intrinsic  AGB stars
  could be as low as 1 M$_\odot$. Some observational evidence for this
  can be found  in the literature \citep{she19}.  This conclusion is
reinforced by  the scale height  onto the  Galactic plane that  can be
estimated for  all the C-rich stars  in Fig. 13 similarly  to what was
done in Sect.  2: an exponential fit gives $z_o\sim  600$ pc, which is
much larger that the scale height derived for the intrinsic N-type AGB
carbon stars that clearly belong to the thin disc (see Sect. 2).

 Finally,  Figure 16  shows the  LF  derived for  these new  candidate
 carbon stars (blue histogram). We selected only objects in the C-rich
 region  in  the  {\it  Gaia}-2MASS diagram,  that  is,  objects  with
 W$_{RP,BP-RP}-$W$_{K_s,J-K_s}< 1.7$  mag, because redder  objects may
 be severely  affected by circumstellar  extinction and the  BC$_K$ is
 rather uncertain  (see Sect.  3). For these  objects, we  estimate an
 uncertainty in  M$_{\rm{bol}}$ of  $\pm 0.30$  mag because  the limit
 adopted for the parallax error is less strict (see above). As Fig. 16
 shows,          the           average          luminosity          is
 $\langle$M$_{\rm{bol}}\rangle=-4.75\pm  0.50$  mag; a  fainter  value
 than that derived for N-type stars in Sect. 3 (see Fig. 7), but still
 in agreement within  the uncertainties. However, the  reason for this
 lower average luminosity  probably is that the selected  stars have a
 lower  metallicity  on average  than  we  studied  in Sect.  3.  This
 metallicity effect is clearly seen in the LF peak of the carbon stars
 in the MCs  \citep[e.g.][]{gul12}, which have a  lower metallicity on
 average  than the  Galactic  counterparts.  Furthermore, this  figure
 shows  a long  tail at  low luminosities  with M$_{\rm{bol}}$  values
 typical of  RGB stars, and  very bright  carbon stars are  also found
 (M$_{\rm bol}<  -5.5$ mag). Figure  16 is  similar to Figure  7. 
   These  luminosity   tails  are  populated  indistinctly   by  stars
   belonging to the thin and thick discs and to halo, according to our
   population   membership   criteria.   This   low-luminosity   tail
 reinforces our  previous conclusion that  many of the  new discovered
 carbon  stars  probably have  an  extrinsic  origin. We  recall  that
 theoretically,  the  minimum luminosity  at  which  the stars  become
 C-rich  during the  AGB  phase is  M$_{\rm{bol}}\sim  -4.0$ mag  (the
 actual   value    depends   on    the   stellar    metallicity;   see
 e.g. \citet{stra06}), and  that the low-luminosity tail in  the LF of
 the N-type  stars (see  Fig. 7,  top panel) can  be explained  with a
 small contribution of these stars (see  also Paper I). When the LF is
 derived only for  stars with luminosites above the  RGB tip (M$_K\leq
 -7.0$ mag; see e.g. \citet{fre20}, the  dotted line in Fig. 16), that
 is,  stars  with   luminosities  typical  of  the   AGB  phase,  this
 low-luminosity tail  disappears and  the minimum in  the LF  is about
 M$_{\rm{bol}}\sim -4$ mag, as expected. Interesting enough, according
 to our  kinematic study, this LF  is built by stars  belonging to the
 different stellar populations, without a significant difference
   among the  individual LF for  each stellar population. If  this is
 confirmed by  more detailed studies,  the existence of  intrinsic AGB
 carbon stars in the thick disc  and halo of the Galaxy might indicate
 that some recent star formation episodes took place in these Galactic
 structures. Another possibility might be that these thick-disc and/or
 halo AGB carbon  stars have an external origin,  perhaps belonging to
 some  of the  stellar  structures with  peculiar kinematics  recently
 discovered in the halo that may  have been captured by our Galaxy and
 that very probably have had  different star formation histories, such
 as the  Gaia-Enceladus-Sausage merger, the Sequoia  or Helmi streams,
 or  the  Thamnos   structure  \citep{Belokurov18,  Helmi18,  Helmi99,
   Myeong19, Koppelman19}.

\begin{figure}
   \centering
   \includegraphics[width=9cm]{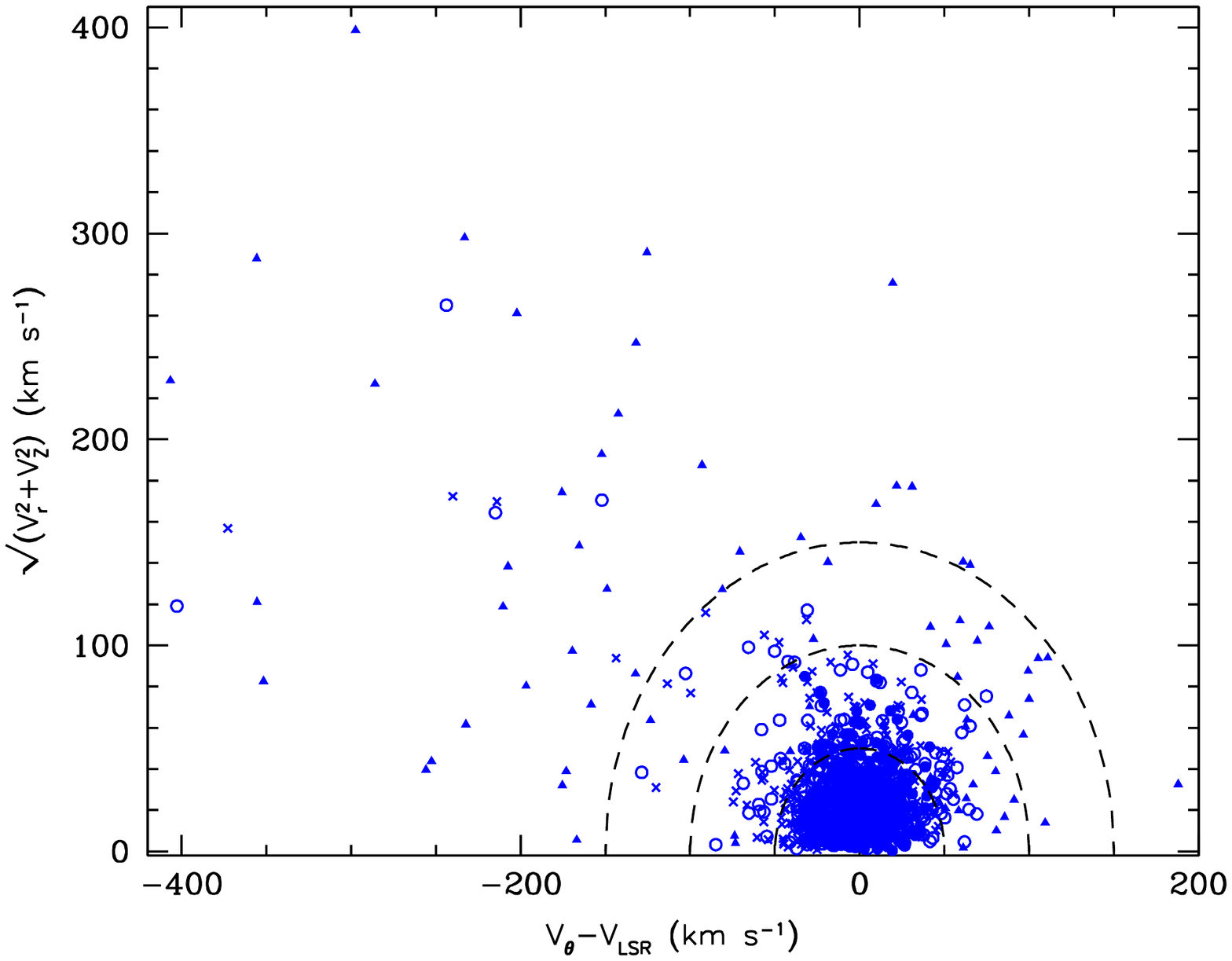}
   \caption{ Toomre diagram for the new candidate carbon stars.
     Stars are plotted according to their membership probability to
     belong to the thin-disc (solid blue circles), thick-disc (crosses),
     or halo (triangles) stellar populations. Open circles refer to stars with
     ambiguous membership (see text). }
  \end{figure}

\begin{figure}
   \centering
   \includegraphics[width=9.0cm]{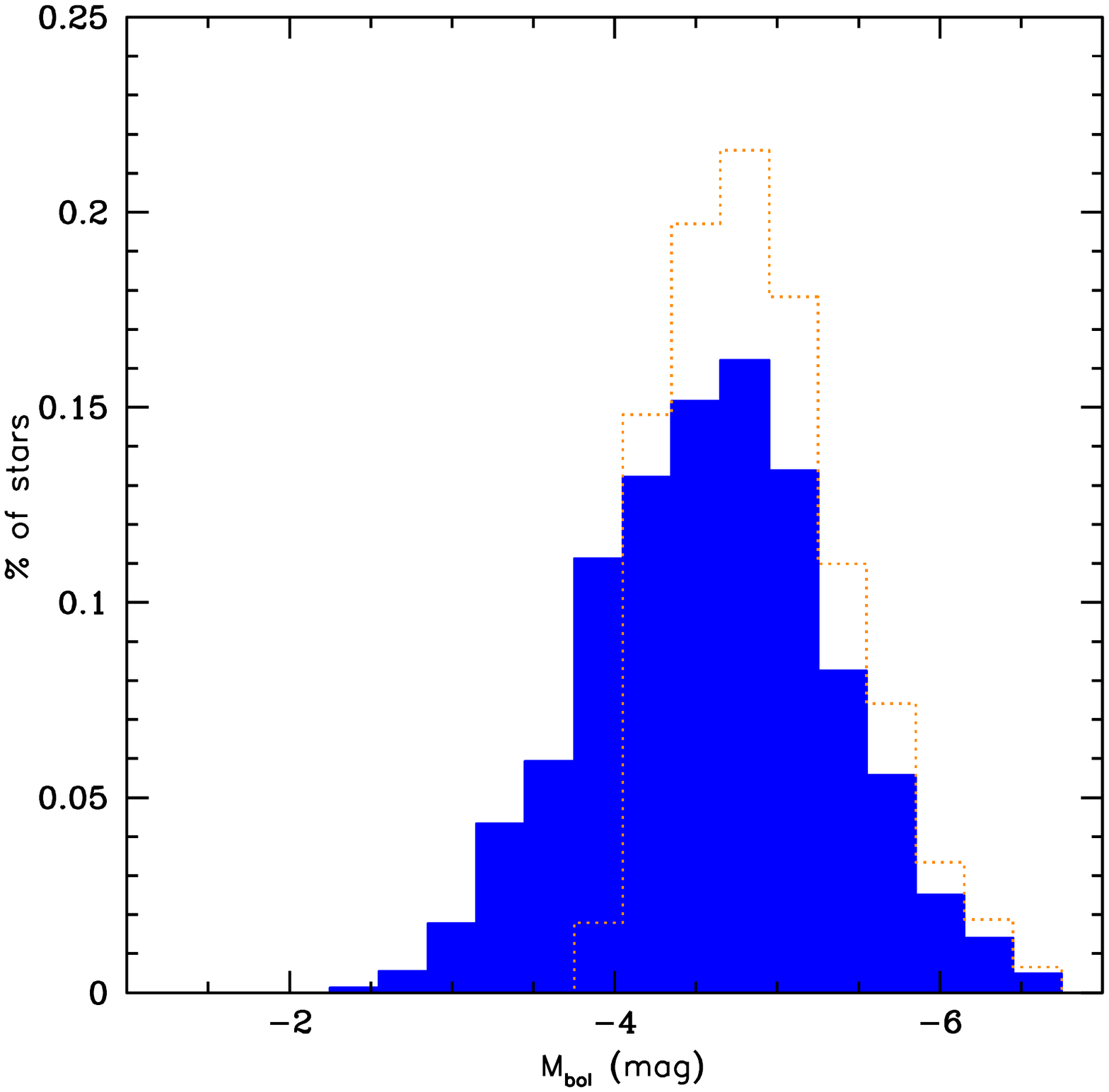}
   \caption{ Luminosity distribution for the new candidate carbon stars (blue)
     with W$_{RP,BP-RP}-$W$_{K_s,J-K_s}< 1.7$ mag. The dotted orange line shows the
     luminosity function only for stars with M$_{K_s}\leq -7.0$ mag, which is
     approximately the luminosity at the RGB tip. The bin size is 0.30 mag.}
  \end{figure}

Obviously, only a detailed  and individual kinematic and spectroscopic
study is able to sort and answer these questions and confirm or reject
the  carbon-rich  nature  of  these candidate  stars.  This  might  be
achieved    with   ground-based    high-resolution   spectra    and/or
space-collected \Gaia/Radial Velocity  Spectrometer spectra that cover
the  Ca  II infrared  triplet  region  at medium  spectral  resolution
\citep{Recio22}.   We  point  out,   however,  that  as  described  by
\citet{mow19}\footnote{See  also,  the  \Gaia\   Image  of  the  Week:
  https://www.cosmos.esa.int/web/gaia/iow-20181115},  the  carbon-rich
nature  of   these  stars   might  also   be  confirmed   through  the
\Gaia\ spectro-photometers, which  provide low-resolution spectra that
allow us to identify possible carbon molecules (C$_2$ and CN) in their
atmosphere.  For these future studies,  the complete list of these new
carbon stars  is available through the  CDS database\footnote{The {\it
    Gaia}-EDR3 identification number of  these stars together with the
  full version of Table 1 are only available in electronic form at the
  CDS via anonymous ftp to cdsarc.u-strasbg.fr (130.79.128.5).}

\section{Summary}
In this study we have extended the previous analysis of \citet[][Paper
  I]{abi20}  to  characterise  Galactic   carbon  stars  of  different
spectral types,  in particular,  to derive their  luminosity function,
location   in  the   Galaxy,   kinematics,   and  stellar   population
membership.  To do  this,  we  made use  of  the accurate  astrometric
measurements  from  {\it Gaia}  EDR3  of  the  stars included  in  the
Galactic  surveys  of  infrared  carbon  stars  by  \citet{che12}  and
\citet{che07},  the LAMOST  survey by  \citet{ji16}, and  the Galactic
S-type stars  by \citet{che19}. From  these surveys, we  selected only
stars  fulfilling  the   criterion  $\epsilon(\varpi)/\varpi\leq  0.1$
having  accurate infrared  photometry. The  final sample  contains 491
carbon stars  of N type,  22 of SC  type, 83 of  J type, 234  of R-hot
type, and 276 of CH type, which  represents an increase by more than a
factor two  in the  number of  objects studied  with respect  to Paper
I.  We  also  added  107  extrinsic and  91  intrinsic  S  stars  from
\citet{che19}.  We  derived  interstellar  extinctions  using  several
Galactic  models  to  test  their impact  on  the  derived  luminosity
functions. Our main results are summarised below.

a) We find that the  interstellar extinction as derived from different
recent Galactic  extinction models  may affect the  luminosity derived
for individual stars,  but has a negligible effect on  the LF function
derived for the different spectral types we studied.

b)   The   average  luminosity   of   carbon   stars  of   N-type   is
$\langle$M$\rm{_{bol}}\rangle =-5.04\pm 0.55$  mag, which confirms our
result  in Paper  I. However,  the derived  LF shows  less significant
tails at  low and high  luminosities, which partially  alleviates some
contradiction with  theoretical predictions  concerning the  lower and
higher   mass  limits   for   a   star  to   become   an  AGB   carbon
star. Nevertheless, a small fraction of N-type stars show luminosities
as high as M$_{\rm{bol}}\sim -6.0$ mag. Although the number of SC-type
stars we analysed  is still small, the LF of  these stars is identical
to that of N-type stars. Our kinematic study also shows that these two
types of carbon stars belong to the thin-disc Galactic population.

c) Our extended  sample of J-type stars confirms that  these stars are
clearly  less  luminous  than  N- and  SC-type  stars.  Their  average
luminosity  is  $\langle$M$_{\rm{bol}}\rangle=-4.42\pm 0.53$  mag.  We
find an  indication of a larger  scale height onto the  Galactic plane
for J-type stars  than for N- and SC-type stars.  We also confirm that
the  LF of  R-hot stars  extends  over a  very wide  range and  covers
luminosities     throughout     the     RGB    (the     average     is
$\langle$M$_{\rm{bol}}\rangle=0.05\pm 1.10$ mag).  This fact precludes
us  from  assigning  these  types   of  carbon  stars  to  a  specific
evolutionary stage, which favours an  extrinsic origin to their carbon
enhancement.   The kinematic  study of  R-hot stars  as well  as their
Galactic location  shows features  that are very  similar to  those of
CH-type stars, which would indicate that these objects might mostly be
old low-mass stars belonging to the Galactic thick disc.

d)  Intrinsic  (O-rich) S  stars  have  higher luminosities  than  the
extrinsic       ones.       Their       average       values       are
$\langle$M$_{\rm{bol}}\rangle=-4.42\pm       0.68$       mag       and
$\langle$M$_{\rm{bol}}\rangle=-3.52\pm  0.68$  mag,  respectively.  In
particular, the luminosities of the overwhelming majority of intrinsic
S stars  are higher  than the  predicted onset of  the TDU  during the
AGB.   For  solar   metallicity  stars,   this  luminosity   limit  is
M$_{\rm{bol}}\sim  -3.5$  mag,  which  corresponds to  stars  with  an
initial mass above 1.3 M$_\odot$. This result is consistent with these
(Tc-rich) stars  being genuine TP-AGB  stars. On the other  hand, both
intrinsic  and  extrinsic S  stars  clearly  belong to  the  thin-disc
population.

e)  We  illustrated again  that  the  2MASS-{\it  Gaia} diagram  is  a
powerful tool  for identifying  C- and O-rich  AGB stars  of different
spectral types, according  to their evolutionary stage  and masses. In
particular, we  find that most of  the carbon stars identified  in the
LAMOST  survey are  not AGB  carbon stars,  but probably  R-hot and/or
CH-type stars.

f) We used the  2MASS-{\it Gaia} diagram to  combine optical
{\it Gaia} and infrared photometry  to identify sub-groups of Galactic
AGB  stars  among  long-period variables,  including  the  distinction
between C-rich  and O-rich  stars. This study  allowed us  to identify
2,659 stars whose  C-rich nature was previously unknown.

g) On the basis of their derived  luminosities and Galactic kinematics,  we argue that
these  new carbon  stars  probably  constitute a  mix  of carbon  star
spectral types  of intrinsic and extrinsic  nature in a wide  range of
metallicities  belonging   to  the  thin-  and   thick-disc  and  halo
populations. The full
list of these new identified carbon stars is available through the CDS
database  for further  studies to  determine their  actual nature  and
evolutionary stage.  An analysis like  this can be performed  based on
\Gaia/RVS      spectra,     ground-based      spectroscopy,     and/or
\Gaia\ low-resolution spectro-photometric data.

\begin{acknowledgements}
This    study    has    been   partially    supported    by    project
PGC2018-095317-B-C21  financed by  the MCIN/AEI  FEDER “Una  manera de
hacer  Europa”.   MRG  and  FF  acknowledge  the  funding  by  Spanish
MICIN/AEI/10.13039/501100011033 and  "ERDF A way of  making Europe" by
the  “European  Union”  through grant  RTI2018-095076-B-C21,  and  the
Institute of Cosmos Sciences University of Barcelona (ICCUB, Unidad de
Excelencia  ’Mar\'{\i}a de  Maeztu’)  through grant  CEX2019-000918-M.
This work  has made use of  data from the European  Space Agency (ESA)
mission  \Gaia\ (\url{https://www.cosmos.esa.int/gaia}),  processed by
the   \Gaia\   Data   Processing  and   Analysis   Consortium   (DPAC,
\url{https://www.cosmos.esa.int/web/gaia/dpac/consortium}).    Funding
for the DPAC has been provided by national institutions, in particular
the   institutions   participating    in   the   \Gaia\   Multilateral
Agreement. This  work has been  partially supported by  the “Programme
National de Physique  Stellaire” (PNPS) of CNRS/INSU  co-funded by CEA
and CNES.  We thank the referee Dr. N. Mowlavi for his useful comments
and suggestions.
\end{acknowledgements}

\bibliographystyle{aa}
\bibliography{main}

\clearpage
\onecolumn

\begin{landscape}
\begin{longtable}{lcccccccccccccc}
\caption{Derived data for the sample stars}\\
\hline
\hline
Name &  \Gaia\ EDR3 source id & $A_V$ & $A_V$ & $A_V$& $J_o$ & $K_{s_o}$ &  distance  & M$_{\rm {bol}}$ &  Z & R$_{GC}$ &  V$_r$ & V$_\theta$  & V$_z$  & Pop  \\
     &                & (mag) & (mag) & (mag) & (mag)     & (mag)           & (pc) &   (mag)               & (kpc) & (kpc) & (km s$^{-1}$)& (km s$^{-1}$)& (km s$^{-1}$)&  \\
     
\hline
N-type & & & & & & & & & & & \\
R Scl & 5016138145186249088      &0.04 &0.08 & ...& 1.43        & $-0.12$       & 387.02 &        $-4.89$ & $-0.36$       & 8.20  &$-60.79$       & 214.43 &       15.14 & 0 \\
AQ Sgr &         4180404407794509056& 0.47&0.26 & 0.45&         2.43&   0.71&   556.30& $-4.88$&        $-0.13$&        7.85&   $-8.70$&        261.10& $-1.70$&        0 \\
TX Psc &  2743004129429424000&0.09 &0.11 &0.07 &        0.79&   $-0.52$&        244.28& $-4.62$&        $-0.18$&        8.19&   $-52.71$&       251.02& $-7.51$&        0 \\
R Lep &  2987082722815713792&0.12 &0.25 & 0.35&         2.11&   0.12&   446.33& $-4.96$&        $-0.21$&        8.50    &$-4.52$&       230.31& 8.82&   0 \\
RT Cap & 6853966780133956096&0.43 &0.28 & 0.45& 2.25&   0.25&   540.51& $-4.92$&        $-0.23$&        7.74&   19.86&  212.20& $-1.21$&        0 \\
RV Cyg &         1952830855365548416&0.46 &0.49 &0.68 & 1.68&   0.69&   670.01& $-6.09$ &       $-0.11$&        8.17&   $-19.86$&       251.43& 7.40    &0 \\
S Sct &  4251993571231284224&0.59 &0.26 &0.24 &         1.86&   0.56&   415.84& $-4.87$&        0.00&   7.81&   $-5.39$&        245.99& $-9.20$&        0\\
SS Vir &         3699854707618766848&0.07 &0.07 & 0.01&         2.32&   0.82&   547.87& $-5.06$&        0.51&   8.10    &$-15.41$       &247.04&        9.44&   0\\
ST Cam &         483958671558728576     & 0.47&0.25 &0.65 & 2.01&       0.42&   586.36& $-5.42$&        0.17&   8.63&   $-7.87$&        243.52& $-7.31$&        0 \\
TU Gem &         3427006189904854656&0.78&0.66 &1.13&   2.20    &0.73   &1101.36&       $-6.44$&        0.09&   9.27&   32.12&  210.41& $-5.98$&        0 \\
... &  & & & & & & & & & & & & & \\
\hline
\end{longtable}

\tablefoot{Pop: membership probability higher than 80\%  for thin disc (0), thick disc (1), halo (2), or ambiguous (3). A blank means that there is no V$_{\rm rad}$ measurement in \Gaia\ DR2 (see text). Columns 3, 4, and 5 show the extinction values from \citet{lal19}, \citet{dri03}, and \citet{gre19} Galactic models, respectively.  The $J_o$ and $K_{s_{o}}$ magnitudes are those corrected for extinction according to \citet{lal19}. The complete version of this table is only available at the CDS.}
\end{landscape}

\end{document}